\documentclass[12pt,journal,draftclsnofoot,onecolumn]{IEEEtran}
\IEEEoverridecommandlockouts

\makeatletter
\def\markboth#1#2{\def\leftmark{\@IEEEcompsoconly{\sffamily}\MakeUppercase{\protect#1}}%
\def\rightmark{\@IEEEcompsoconly{\sffamily}\MakeUppercase{\protect#2}}}
\makeatother

\usepackage[english]{babel}
\usepackage{color}
\usepackage{lipsum}
\usepackage{caption}
\usepackage{cite}
\usepackage[pdftex]{graphicx}
\usepackage[font=footnotesize]{subcaption}
\usepackage{amsmath}
\usepackage{amsfonts}
\usepackage{array}
\usepackage{verbatim}
\usepackage{listings}
\usepackage{algorithm}
\usepackage{algpseudocode}
\usepackage{hyperref}
\usepackage{url}
\usepackage{enumerate}
\usepackage{multirow}

\usepackage{epsfig}
\usepackage{epstopdf}
\usepackage{multicol}
\usepackage[font=footnotesize]{caption}

\DeclareMathOperator*{\argmax}{arg\,max}
\renewcommand{\arraystretch}{2}

\newcommand{\bi}{\begin{itemize}}
\newcommand{\ei}{\end{itemize}}
\newcommand{\be}{\begin{equation}}
\newcommand{\ee}{\end{equation}}

\addto\captionsenglish{}

\addto\captionsenglish{}

\def\beq{\begin{equation}}
\def\eeq{\end{equation}}
\def\beqa{\begin{eqnarray}}
\def\eeqa{\end{eqnarray}}
\def\beqan{\begin{eqnarray*}}
\def\eeqan{\end{eqnarray*}}

\usepackage{makecell}

\usepackage{adjustbox}

\newcommand{\captionaboveof}[3][]{%
    \vskip-\abovecaptionskip
    \vskip+\belowcaptionskip
    \def\@captype{#2}%
    \ifx\@nnil#1\@nnil
        \caption{#3}%
    \else
        \caption[#1]{#3}%
    \fi
    \vskip+\abovecaptionskip
    \vskip-\belowcaptionskip
}
\makeatother

\usepackage{threeparttable}
\usepackage[table,xcdraw]{xcolor}
\usepackage{tabularx}
   \usepackage{multirow}
   \usepackage{booktabs}
\newcommand{\tabitem}{~~\llap{\textbullet}~~}
   \usepackage{graphicx}
   \usepackage{array, blindtext}
   \usepackage{wrapfig}
\usepackage{pdfpages}

\usepackage{tikz}
\usepackage{pgfplots}
\pgfplotsset{compat=newest} 
\pgfplotsset{plot coordinates/math parser=false} 
\newlength\fheight
\newlength\fwidth
\usetikzlibrary{patterns,decorations.pathreplacing,backgrounds,calc,spy,arrows.meta}
\definecolor{SchoolColor}{RGB}{0.71, 0, 0.106}
\definecolor{chaptergrey}{rgb}{0.61, 0, 0.09} 
\definecolor{midgrey}{rgb}{0.4, 0.4, 0.4}
\definecolor{chaptergreen}{rgb}{0.09, 0.612, 0}
\definecolor{chapterpurple}{rgb}{0.522, 0, 0.612}
\definecolor{chapterlightgreen}{rgb}{0, 0.612, 0.522}

\newcommand*\circled[1]{\tikz[baseline=(char.base)]{
            \node[shape=circle,draw,inner sep=4pt] (char) {#1};}}
            
            \usepgfplotslibrary{colorbrewer}

\title{An \hspace*{-0.1cm} Efficient \hspace*{-0.3cm} Uplink \hspace*{-0.3cm}  Multi-Connectivity  \hspace*{-0.3cm} Scheme  \hspace*{-0.3cm} for \hspace*{-0.3cm}   5G \hspace*{-0.3cm}   mmWave  \hspace*{-0.3cm}  Control \hspace*{-0.3cm} Plane \hspace*{-0.3cm} Applications }

\author{{{ Marco Giordani}$^\dagger$, { Marco Mezzavilla}$^\diamond$, { Sundeep Rangan}$^\diamond$, { Michele Zorzi}$^\dagger$ }\\
\normalsize $^\dagger$ Department of Information Engineering (DEI), University of Padova, Italy \\
\normalsize $^\diamond$NYU Wireless, Brooklyn, NY, USA \\
\small{$\{$\texttt{giordani}, \texttt{zorzi}$\}$\texttt{@dei.unipd.it}, $\{$\texttt{mezzavilla}, \texttt{srangan}$\}$\texttt{@nyu.edu}
}
\thanks{A preliminary version of this paper was
presented at the  \emph{15th Annual Mediterranean Ad Hoc Networking Workshop (Med-Hoc-Net)}, Vilanova i la Geltru, Barcelona, Spain, June 2016 \cite{MedHoc2016}.}}

\begin{document}
\maketitle

\tikzstyle{startstop} = [rectangle, rounded corners, minimum width=2cm, minimum height=0.5cm,text centered, draw=black]
\tikzstyle{io} = [trapezium, trapezium left angle=70, trapezium right angle=110, minimum width=3cm, minimum height=1cm, text centered, draw=black]
\tikzstyle{process} = [rectangle, minimum width=2cm, minimum height=0.5cm, text centered, draw=black, align=center]
\tikzstyle{decision} = [ellipse, minimum width=2cm, minimum height=1cm, text centered, draw=black]
\tikzstyle{arrow} = [thick,<->,>=stealth]
\tikzstyle{line} = [thick,>=stealth]
\tikzstyle{darrow} = [thick,<->,>=stealth,dashed]
\tikzstyle{sarrow} = [thick,->,>=stealth]

\begin{abstract}  The millimeter wave (mmWave) frequencies offer the potential of orders of magnitude
increases in capacity for next-generation cellular  systems.  However, links in mmWave networks are 
susceptible to blockage and may suffer from rapid variations in quality.
Connectivity to multiple cells --  at mmWave and/or traditional  frequencies -- is  considered
essential for robust communication.
One of the challenges in supporting multi-connectivity in  mmWaves
 is the requirement for the network  to track the direction of each link in addition to its
power and timing.
To address this challenge, we
implement a novel uplink measurement system that, with the joint help of a local coordinator operating in the legacy band, guarantees continuous monitoring of the channel propagation conditions and allows for the design of efficient control plane applications, including handover, beam tracking and initial access.
We show that an uplink-based multi-connectivity approach enables less consuming, better performing, faster and more stable cell selection and scheduling decisions with respect to a traditional downlink-based standalone scheme. 
Moreover, we argue that the presented framework guarantees (i) efficient tracking of the user in the presence of the channel dynamics expected at mmWaves, and (ii) fast reaction to situations in which the primary propagation path is blocked or not available.
\end{abstract}

\begin{IEEEkeywords}
5G, millimeter wave, multi-connectivity, initial access, handover, blockage, beam tracking.
\end{IEEEkeywords}

\section{Introduction}

The millimeter wave (mmWave) bands -- roughly above $10$~GHz --
have attracted considerable attention
for meeting the ever more demanding performance requirements of
micro and picocellular networks \cite{RanRapE:14}.
These frequencies offer much more
bandwidth than current cellular systems in the congested  bands below 6~GHz, and initial capacity estimates have suggested that mmWave networks  can offer orders of magnitude higher bit-rates than 4G systems \cite{Mustafa}.

However, the increased carrier frequency of mmWave systems makes the propagation conditions more demanding than at the lower frequencies traditionally used for wireless services, especially in terms of robustness.
MmWave signals are blocked by many common building materials such as brick, and  the human body can also
significantly attenuate signals in the mmWave range \cite{lu2012modeling}.  Thus, the communication quality
between the user equipment (UE)
and any one cell can be highly variable as the movement of obstacles or even the changing position of the
body relative to the mobile device can lead to rapid drops in signal strength.
One likely key feature of mmWave cellular networks that can improve robustness is
\emph{ multi-connectivity} (MC) \cite{MC}, which enables each UE to maintain multiple possible signal paths to different
cells so that drops in one link can be overcome by switching data paths.
Multi-connectivity can be  both among multiple 5G mmWave cells
and between 5G mmWave cells and traditional 4G cells below 6~GHz.
Mobiles with such 4G/5G multi-connectivity feature can benefit from both the high bit-rates that can be provided by
the mmWave links, as well as the 
more robust, but lower-rate, legacy channels,
 thereby opening up new ways of solving capacity issues, as well as new ways of  providing good mobile network performance and robustness \cite{ericsson_1}.

This paper addresses one of the key challenges in supporting multi-connectivity in
heterogeneous networks (HetNets) with mmWave cells, namely directional multi-cell channel tracking
and measurement reports.
These operations are fundamental for cellular systems
to properly perform a wide variety of control tasks including handover,
path selection, and radio-link failure (RLF) detection and recovery.
However, while channel tracking and reporting is relatively straightforward
in cellular systems at conventional frequencies, the mmWave bands present
several significant limitations, including:
(i)~the high variability of the channel in each link due to blockage \cite{George_ICC, barati2015initial};
(ii) the need to track multiple directions for each link \cite{EW2016}; and
(iii) reports from the UE back to the cells must be made directional \cite{magazine_IA}.
\subsection{Contributions}

To address these challenges, in this paper we provide the first comprehensive numerical evaluation of the performance of a novel uplink (UL) multi-connectivity
measurement reporting system which enables fast, robust and efficient cell selection.
In such a scheme, the UE directionally broadcasts sounding reference signals (SRSs) in time-varying directions that continuously
 sweep the angular space. Each potential serving cell scans all its angular directions
 and monitors the strength of the received SRSs.
 A centralized controller (that can be identified by an LTE eNB operating in the legacy band)  obtains complete directional knowledge from all the potential cells
 in the network to make the optimal serving cell selection and
 scheduling decision. 
We note that the proposed scheme should not be confused with a mmWave version of coordinated multipoint (CoMP) 	\cite{Irmer2011}.  In CoMP, multiple eNBs simultaneously transmit to obtain beamforming gains across cells (in effect creating a high-dimensional antenna array).  In the proposed method, although it measures control signals from multiple cells, the UE receives data from only one cell at a time.  Importantly, unlike traditional CoMP, the eNB does not need to maintain relative phase information for the links from different cells - a task that would be extremely difficult in the mmWave setting due to the high Doppler.  The proposed method is thus closer to carrier aggregation or fast handover.
 
 As an extension of our work \cite{MedHoc2016}, in this paper we additionally aim at  comparing the performance of the presented approach with respect to a traditional downlink-based standalone (SA) scheme. We numerically show that:

\begin{itemize}
\item The implementation of an UL-based framework enables a faster and less energy consuming tracking of the channel quality over time at the mobile terminal. In fact, an uplink sounding scheme  eliminates the need for the UE
to send measurement reports back to the network and thereby removes a possible point of failure in the control
signaling path. Moreover, if digital beamforming or beamforming with multiple analog streams
is available at the mmWave cell, then the
directional scan time can be dramatically reduced when using UL-based
measurements.
\item The use of a MC approach enables a better performing resource allocation and mobility management with respect to a SA configuration. 
In fact, the LTE connectivity can offer a ready backup in case the mmWave links suffer an outage and can be used to  forward the scheduling and serving  decisions to the  user if the main propagation path in unavailable.
\item The presented framework guarantees robust and stable communication quality  in the presence of the channel variations and dynamics expected at mmWaves.
\end{itemize}

Furthermore, we give numerical evidence  of how the proposed UL-based framework enables the design of efficient 5G control plane applications and fundamental MAC layer functions that specify how a UE should connect to the network and preserve its connectivity. Specifically, our scheme allows for: 
\begin{itemize}
\item Efficient and stable \emph{handover}. Dense deployments of short range cells, as foreseen in future mmWave cellular networks, may exacerbate frequent handovers between adjacent eNBs \cite{zorzi}. High throughput values can be continuously guaranteed when intensively monitoring the UE's channel quality over time (even when considering highly dynamic environments).
\item  Fast and fair \emph{initial access}. Unlike in traditional attachment policies, by leveraging on the presence of the local coordinator, the initial association can be possibly performed by taking into account the instantaneous load conditions of the surrounding cells, thereby promoting fairness in the whole cellular network.
\item Reactive \emph{RLF detection and recovery}. By exploiting previously saved instances of the channel quality information exchanged by the network nodes, a backup steering direction can be set  in case the primary one is blocked, to immediately recover an acceptable communication service without waiting for a handover to be eventually triggered.
\end{itemize}

Finally, we evaluate the performance of the presented framework by considering a detailed real-world measurement-based mmWave channel scenario, for which we defined an innovative mobility model which accounts for the dynamics (in terms of both small and large scale fading) experienced by the mmWave links. 
Most of the studies so far have been conducted in stationary conditions with minimal local blockage, whereas this is one of the first contributions in which a dynamic environment is considered.

\subsection{Related Work}
\label{sec:rel_work}

Channel estimation is relatively straightforward in  LTE \cite{schwarz2010calculation}. However, in addition to the rapid variations of the channel, transmissions in mmWaves  are expected to be  directional, and thus
 the network and the UE must constantly monitor the direction
of transmission of each potential link.
Tracking changing directions can slow the rate at which
the network can adapt, and can be a major obstacle in providing robust service in the face of variable link
quality.  In addition, the UE and the eNB  may only be able to listen to one direction at a time,
thus making it hard to receive the control signaling necessary to switch paths.

Dual-connectivity has been proposed in Release $12$ of Long Term Evolution-Advanced (LTE-A) \cite{3GPP_DC}.  This feature supports inter-frequency and intra-frequency connectivity
as well as connectivity to different types of base stations (e.g., macro and pico base stations) \cite{sota_MC2}.  However, these systems were designed for conventional 
frequencies, and did not address the directionality and variability of the channels present at mmWave
frequencies.

Some other previous works, such as \cite{sota_MC},   
consider the bands under $6$ GHz as the only control channel for 5G networks, to provide robustness against blockage and wider coverage range. 
However, high capacities can also be obtained just exploiting the mmWave frequencies. So, in \cite{MC}, a multi-connectivity framework is proposed as a solution for mobility-related link failures and throughput degradation of cell-edge users, relying  on the fact that the transmissions from cooperating cells are coordinated for both data and control signals.

The work in \cite{sota_MC3}  assumes a HetNet deployment of small cells and proposes that the control plane be handled centrally for small geographical areas whereas, for large geographical areas,  distributed  control should be used. However, the  performance evaluation of  small cells that use the same carrier frequency deployed over a relatively wider area has not  yet been investigated.


Finally, in \cite{JSAC_2017} we showed, through an extensive simulation campaign, that the proposed framework is suitable to enable fast network handover procedures. However, we did not investigate the performance of other interesting cellular control applications (i.e., initial access or RLF detection and recovery) and we did not account for the dynamics that affect the mmWave~propagation.

\section{Uplink Multi-Connectivity Procedure Description}
\label{sec:MCP}


In the presented framework, summarized  in Tab. \ref{tab:procedure}, there is one major node called MCell (Master Cell, in accordance with  3GPP LTE terminology), which here is typically
an LTE eNB operating in the legacy band. However, functionally, the MCell can be any network entity
that performs centralized handover and scheduling decisions.
The UE may receive data from a number of mmWave cells we refer to as SCells (Secondary Cells).
In order to communicate and exchange control information, the S/MCells are interconnected via traditional backhaul X2  connections, while each user can be reached by its serving MCell through the legacy~band.

MmWave SCells and UEs will likely utilize directional phase arrays for transmission.
In this work, we  assume that nodes select one of a finite number of directions,
and we let $N_{\rm SCell}$ and $N_{\rm UE}$ be the number of directions at each SCell
and UE, respectively.  Thus, between any cell and the UE there are a total of
$N_{\rm SCell} \times N_{\rm UE}$ direction pairs.  The key challenge in implementing
multi-cell connectivity is that the network must monitor the signal strength
on each of the direction pairs for each of the possible links. 
This is done by each SCell  building a \emph{report table} (RT) for each user, based on the channel quality of each receiving direction  that can be used by the central entity to: 
(i) help the UE  identify the  mmWave eNB with the best instantaneous propagation conditions, and (ii)  trace and estimate, over time, the channel quality conditions.

The system can be more precisely described as follows.
Suppose that, in the considered area, $M$  SCells and $N$ UEs are deployed under the control of one MCell.
The framework
performs its monitoring operations through three main phases.

\renewcommand{\arraystretch}{0.999}
\begin{table}[!t]
\centering
\caption{Description of the uplink-based multi-connectivity procedure presented in Sec. \ref{sec:MCP}}.
\begin{tabular}{|c|c|}
\hline
\makecell{ \textbf{First Phase} \\ -- UL Measurements --} & 
\thead{
\tabitem UE transmits the SRSs to the surrounding SCells, through directions $d_1, \dots, d_{N_{\rm UE}}$. \\
\tabitem SCell performs an exhaustive search to collect the SRSs, through directions $D_1, \dots, D_{N_{\rm SCell}}$.\\
\tabitem RTs are filled with the metrics SINR$_{i,j}$, $\forall i\in \{1,\dots,N\}$, $\forall j\in \{1,\dots,M\}$.
}
\\
\hline
\makecell{\textbf{Second Phase} \\ -- Coordination -- } & 
\thead{
\tabitem SCell sends its RT to the MCell, through the backhaul link. \\
\tabitem MCell builds a CRT by collecting all the received RT.\\
\tabitem MCell makes  attachment decisions by selecting the optimal SCell for each UE to connect to. $\quad$
}
\\
\hline
\makecell{\textbf{Third Phase} \\ -- Decision --} & 
\thead{
\tabitem  MCell forwards the best decision for the transceiver. \\
\tabitem SCell is informed through the backhaul link. \\
\tabitem UE is informed through the LTE legacy band, to remove a  point of failure in the feedback.
}\\
\hline
\end{tabular}
\label{tab:procedure}
\end{table}


\subsection{First phase: Uplink measurements}

In the MC procedure's first phase, each SCell fills  its RT.
Each UE directionally broadcasts uplink sounding reference signals in dedicated slots, steering through directions $d_1,\dots,d_{N_{\rm UE}}$, one  at a time, to cover the whole angular space.
The SRSs are scrambled by locally unique identifiers (e.g., C-RNTI) that are known to the
SCells.
We are therefore exploiting an UL measurement reporting system where, unlike in traditional mechanisms,  the reference signals are broadcast by each UE rather than by the eNBs. The advantages of this design choice will be explained in the next sections of this work.
If analog beamforming is used, each SCell performs an exhaustive search, scanning through directions $D_1,\dots,D_{N_{\rm SCell}}$, one at a time or, if digital beamforming is applied, from all of them at  once\footnote{The synchronization between the sweeping of the users and the listening of the base stations in the mmWave  band is guaranteed by assuming  that the mobile terminals have already exchanged some preliminary time/frequency synchronization information through the LTE connectivity.}.

Each SCell fills its RT whose entries represent the highest Signal-to-Interference-plus-Noise-Ratio (SINR) between UE$_i$, $i\in \{1,\dots,N\}$, transmitting through its best direction $d_{\rm UE,opt} \in \{d_1,\dots,d_{N_{\rm UE}}\}$ and the SCell$_j$, $j \in \{1,\dots M\}$, receiving through
its best possible direction  $D_{ \rm SCell,opt} \in \{D_1,\dots,D_{N_{\rm SCell}}\}$:
\beq
\text{SINR}_{i,j}(d_{\rm UE,opt},D_{\rm SCell,opt}) = \argmax_{\substack{d_{\rm UE}=d_1,\dots,d_{N_{\rm UE}}\\ D_{\rm SCell} =D_1,\dots,D_{N_{\rm SCell}}}}  \text{SINR}_{i,j}(d_{\rm UE},D_{\rm SCell})
\label{eq:max_SINR}
\eeq

\subsection{Second Phase:  Coordination}

Once the RT of each SCell has been filled, each mmWave cell sends
this information, through the backhaul link, to the supervising MCell which, in turn, builds a \emph{complete report table (CRT)}, as depicted in Tab. \ref{tab:RT}. The controller has indeed a complete overview of the surrounding channel conditions and gains a comprehensive vision over the whole cellular system it oversees.
When accessing the CRT, the MCell eventually makes a network decision by  selecting the best candidate mmWave SCell for each user to connect to, based on different metrics (i.e., the maximum SINR, with some hysteresis, or the maximum rate, when being aware of the current load of each cell).

For example, if the maximum SINR attachment policy is selected,  the UE$_i$, $i\in \{1,\dots,N\}$ will be associated with the mmWave SCell$_j$, $j\in \{1,\dots,M\}$, if the entry 
$${\rm SINR}_{i,j}(d_{\rm UE,opt},D_{\rm SCell,opt})$$
 is the highest one in the $i$-th row of the CRT.
Such maximum SINR is associated, in the CRT's entry, to the SCell (UE) optimal direction $D_{\rm SCell, opt} $ ($d_{\rm UE, opt} $), which should therefore be selected to reach the UE (SCell) with the best  performance.

\begin{table}[!t]
\centering
\caption{An example of CRT, referred to $N$ users and $M$ available mmWave SCells in the network. We suppose that the UE can send the sounding signals through $N_{\rm UE}$ angular directions and each mmWave eNB  can receive them through $N_{\rm SCell}$ angular directions. Each pair is the maximum SINR measured in the best direction between the UE ($d_{\rm UE,opt}$) and the SCell ($D_{\rm SCell,opt}$).}
\renewcommand{\arraystretch}{0.9}
\begin{adjustbox}{max width=\columnwidth}
\begin{tabularx}{0.7\textwidth}{ @{\extracolsep{\fill}} cccc}
\toprule
\multicolumn{4}{c @{\extracolsep{\fill}}}{Complete Report Table (CRT)} \\
\midrule
UE & mmWave SCell$_1$ & \dots & mmWave SCell$_M$\\
\midrule
UE$_1$ & SINR$_{1,1}(d_{\rm UE,opt},D_{\rm SCell,opt})$ &  \dots & SINR$_{1,M}(d_{\rm UE,opt},D_{\rm SCell,opt})$ \\
UE$_2$ & SINR$_{2,1}(d_{\rm UE,opt},D_{\rm SCell,opt})$ &  \dots & SINR$_{2,M}(d_{\rm UE,opt},D_{\rm SCell,opt})$ \\
\dots &\dots  &\dots &\dots \\
UE$_N$ & SINR$_{N,1}(d_{\rm UE,opt},D_{\rm SCell,opt})$ &  \dots & SINR$_{N,M}(d_{\rm UE,opt},D_{\rm SCell,opt})$ \\
\bottomrule
\end{tabularx}
\end{adjustbox}
\label{tab:RT}
\end{table}

\subsection{Third Phase: Network Decision}
If the serving cell needs to be switched, or a secondary cell needs to be added or dropped,
 the MCell needs to inform both the UE and the mmWave eNB.
Since the UE may not be listening in the direction of the target SCell,
the UE may not be able to hear a command from that cell.  
Moreover, since path switches and cell additions in the mmWave regime
are commonly due to link failures, the control link to the serving mmWave cell may not be available either.
To handle these circumstances, we propose that the path switch and scheduling commands
be communicated over the coordinator operating in the legacy band.

Therefore, the MCell notifies  the designated optimal mmWave SCell, via the high capacity backhaul,  about the UE's desire to attach to it. It also embeds the best direction $D_{\rm SCell,opt}$ that should be set to reach that user.
Moreover,
supposing that the UE has already set up a link to the LTE eNB, on a legacy connection,
 the MCell sends to the UE, through an omnidirectional control signal at sub-6 GHz frequencies, the best user direction $d_{\rm UE,opt}$ to select, to reach such candidate SCell.
 By this time, the best SCell-UE beam pair has been determined, therefore the transceiver can directionally communicate in the mmWave band.

 We recall that the attachment decision is  performed neither  by the user nor by the designated mmWave SCells, but rather by the supervising MCell, which is the only entity having a clear and complete  overview of the channel propagation conditions. This guarantees much higher reliability (since the low frequencies at which the coordinator operates can easily penetrate through obstacles)  and fairness (since the attachment decision is periodically triggered by taking into consideration the propagation conditions of the whole cellular network) in the communication~system.

%
%

\section{Enabling 5G Control Applications}
\label{sec:hints}

Existing MAC and lower-layer control procedures  already implemented in a variety of traditional wireless systems should be revised and adapted to the unique mmWave radio environment in which next-generation networks are expected to operate.

\textbf{Cellular.} Next-generation cellular
systems must provide a mechanism by which UEs and mmWave eNBs establish highly directional transmission links, typically formed with high-dimensional phased
arrays, to benefit from the resulting beamforming gain
and balance for the increased isotropic pathloss experienced at high frequencies. In
this context, directional links require fine alignment of the
transmitter and  receiver beams, an operation which might
dramatically increase the time it takes to access the network \cite{magazine_IA}. Moreover, the dynamics of the mmWave channel imply
that the directional path to any cell can deteriorate rapidly,
necessitating  an intensive tracking of the mobile
terminal.

\textbf{Vehicular.} Advanced and sophisticated sensors future cars will be equipped with will require an unprecedented amount of data to be exchanged, which goes beyond the capabilities of existing technologies. 
On the one hand, the mmWave frequencies have the potential to support the required higher
data rates.
On the other hand, there are many concerns regarding the transmission
characteristics of the mmWave channels in an automotive environment \cite{MOCAST_2017}.
For instance, once the nodes are directionally connected, the alignment
can be maintained by using beam tracking mechanisms that try
to maintain a consistent view of the most favorable beam directions over time. However, in highly dense or highly mobile
vehicular scenarios, the corresponding peer may change frequently and may  not last long enough to allow the completion of a data exchange, thus resulting
in transmission errors.
 Moreover, the increased Doppler effect
could make the assumption of channel reciprocity not valid
and could impair the feedback over mmWave links, which
is a potential point of failure for beam sweeping.
Re-alignment of the beams is therefore required to maintain
connectivity.

\textbf{802.11ad.}
The IEEE 802.11ad standard operates in the 60 GHz millimeter wave spectrum and therefore currently designed control protocols already address some of the requirements and challenges pertaining to a high-frequency environment.
However, most proposed solutions are unsuitable for future
5G mmWave mobile network requirements, since  they
present many limitations (e.g., they are appropriate for short-range, static, and indoor scenarios,
which do not match well the requirements of 5G systems). Therefore, new specifically designed
solutions for dynamic networks need to be found.

As we will numerically show in Sec. \ref{sec:res} and discuss in Sec. \ref{sec:HO} -- \ref{sec:RLF}, we claim that faster, more efficient and more robust control plane applications (including handover, beam tracking, initial access, RLF  recovery) can be enabled  when considering a multi-connectivity system, with respect to the case in which a standalone scheme is preferred.

\subsection{Handover and Beam Tracking}
\label{sec:HO}

Handover  is performed when the UE moves from the coverage of one cell to the coverage of another cell \cite{LTE_book}. Beam tracking refers to the need for a user to periodically adapt its steering direction, to realign with its serving eNB, if it has moved or the channel propagation conditions have changed over time.
 Frequent handover, even for fixed UEs, is a potential drawback of mmWave systems due to their vulnerability to random obstacles, which is not the case in LTE. Dense deployments of short range eNBs, as foreseen in mmWave  networks, may exacerbate frequent handovers between adjacent eNBs. Loss of beamforming information due to channel change is another reason for handover and reassociation~\cite{zorzi}.

The literature on handover and beam tracking in more traditional sub-6 GHz heterogeneous networks is quite mature.
For instance, the survey in \cite{Yan20101848} and the work in \cite{HO_WiFi} present multiple vertical handover decision algorithms that
are essential for  wireless networks while, in \cite{schwarz2010calculation}, omnidirectional pilots are used for  channel estimation. However,  most works are specifically tailored to low-frequency legacy cellular systems, whose features are largely different from those of a mmWave environment, preventing the proposed techniques from being applicable to next-generation 5G scenarios.
On the other hand, papers on mobility management for mmWave networks (e.g., \cite{handoff,HO_train,HO_het,HO_het_2,JSAC_2017}) are very recent,
since research in this field is just at its infancy.

We argue that, when considering a higher frequency setting, the presented MC framework can ensure more efficient mobility management operations by exploiting the centralized MCell control over the network to: (i) periodically determine the UE's optimal  mmWave SCell to associate with (if  handover  is strictly required\footnote{In order to reduce the handover frequency, more sophisticated decision criteria could be investigated, rather than triggering a handover every time a more suitable SCell is identified (i.e., the reassociation might be performed only if the SINR increases above a predefined threshold, with respect to the previous time instant). A more detailed discussion of the different handover paradigms is beyond the scope of this paper and we refer the interested reader to \cite{JSAC_2017} for further details.}), or (ii) the new direction through which it should steer the beam (if a simple beam adaptation event guarantees a sufficiently good communication quality), when the user is in \emph{connected-mode}, i.e., it is already synchronized with both the LTE and the mmWave cells. The key input information for the handover/beam tracking decision includes (i) instantaneous channel quality, (ii)  channel robustness, and (iii) cell occupancy.

With respect to the existing algorithms, the use of both the sub-6 GHz and the mmWave control planes is a key functionality for such a  technique. In fact, especially when considering highly unstable and scarcely dense scenarios, the LTE connectivity ensures a ready backup in case the mmWave links suffer an outage. 
Furthermore, the handover/beam switch decision is forwarded to the UE through the MCell, whose legacy link is much more robust and less volatile than its mmWave counterpart, thereby removing a possible point of failure in the control signaling path. Since each SCell periodically forwards the RT, the MCell has a complete overview of the cell dynamics and propagation conditions and can accordingly make network decisions, to maximize the overall performance of the cell it oversees, as we will numerically show in Sec. \ref{sec:HO_res}.
Moreover, unlike in the traditional procedures in which  the users are not aware of the  surrounding cells' current state, 
the UE may choose to connect to the SCell providing either the maximum SINR or the maximum rate (depending on what is considered more convenient), thus taking into account the load of each mmWave cell.

We finally remark that, if previous versions of the report table are kept as a record, the  MCell can also use the SCells' quality variance in selecting the mmWave eNB a user should attach to, after a handover is triggered. If a selected SCell shows a large variance (which reflects high channel instability), the user might need to handover again in the very near future. Therefore, it could be better to trigger a handover only to  an SCell which grants both good SINR (or rate) and sufficient channel stability, leading to a more continuous and longer-term network association. We will address this analysis as part of our future work.

\subsection{Initial Access}
\label{sec:IA}
 The  procedure described in the previous subsection is referred to a UE that is already connected to the network.
However, we show that an uplink-based multi-connectivity control approach may  allow for fast initial access (IA)
from idle mode too.
Initial access \cite{CISS, magazine_IA}  is the procedure by which a mobile UE establishes an initial
physical link connection with a cell, a necessary step to access the network.
In current LTE systems, IA is performed on omnidirectional channels \cite{LTE_book}.
However, in mmWave cellular systems, transmissions will need to be directional
to overcome the increased isotropic pathloss experienced at higher frequencies.
IA must thus
provide a mechanism by which the eNB and the UE can determine suitable initial directions of transmission.

MmWave initial access  procedures have been recently analyzed in \cite{Barati,barati2015initial,Capone,alk}.
Different design options have been compared in \cite{CISS,magazine_IA}, to evaluate coverage and access delay. We refer to \cite{magazine_IA} for a more detailed survey of recent IA works.
All of these methods are based on the current LTE design in which each cell broadcasts
synchronization signals and each UE scans the directional space to  find the optimal node to potentially connect to.
A key result of these schemes is that the dominant delay in downlink-based IA arises in this initial
sychronization phase.
We therefore propose an alternate uplink scheme, as shown in Fig. \ref{fig:initial_access_procedure_fig}, primarily based on the MC framework described in Sec. \ref{sec:MCP}. According to the LTE terminology:

\begin{figure}[!t]
\centering
  \begin{tikzpicture}[font=\sffamily\small, scale=0.6, every node/.style={scale=0.6}]
    \coordinate (UeBegin) at (0,10);
    \coordinate (UeEnd) at (0,0);
    \coordinate (mmWaveEnb1Begin) at (10,10);
    \coordinate (mmWaveEnb1End) at (10,0);
    \coordinate (mmWaveEnb2Begin) at (20,10);
    \coordinate (mmWaveEnb2End) at (20,0);

    \node [above of=UeBegin, yshift=-15pt] (UElabel) {\normalsize SCell};
    \node [above of=mmWaveEnb1Begin,yshift=-15pt] (mmWaveEnbLabel) {\normalsize UE};
    \node [above of=mmWaveEnb2Begin,yshift=-15pt] (mmWaveEnbLabel) {\normalsize MCell};

    \begin{scope}[on background layer]
      \draw [thick, black] (UeBegin) -- (UeEnd);
      \draw [thick, black] (mmWaveEnb1Begin) -- (mmWaveEnb1End);
      \draw [thick, black] (mmWaveEnb2Begin) -- (mmWaveEnb2End);
    \end{scope}

    \filldraw[fill=midgrey!20] (9.8, 9.5) rectangle (20.2,8.5);
    \node (initialLte) at (15,9) { $\circled{0}$ \: LTE Initial Access };

    \draw [-{Latex}] (10, 8.3) -- node[sloped, anchor=center, above ] {$\circled{1}$ Send periodic RAP} (0, 8);
       \draw [-{Latex}] (10, 7.7) -- node[sloped, anchor=center, above] {\vdots} (0, 7.4);
          \draw [-{Latex}] (10, 7.5) -- node[sloped, anchor=center, above] {} (0, 7.2);
    \draw [-{Latex}, red,thick,dashed] (0, 6.5) -- node[sloped,  above, xshift=-5cm] {$\circled{2}$ Send RT} (20, 6.1);

    \draw [-{Latex}, red,thick,dashed] (20, 5.4) -- node[sloped,  above, xshift=5cm] {$\circled{3a}$ Forward optimal direction ($D_{\rm SCell,opt}$)} (0, 5);
    
    \draw [-{Latex}, black!40!green,thick,dash dot] (20, 4.5) -- node[sloped,  above] {$\circled{3b}$ Forward optimal direction ($d_{\rm UE,opt}$)} (10, 4.1);
    
        \draw [-{Latex}] (0, 3.4) -- node[sloped, anchor=center, above ] {$\circled{4}$ RAR} (10, 3);
        
        \draw [-{Latex}] (10, 2.3) -- node[sloped, anchor=center, above ] {$\circled{5}$ CRM} (0, 1.9);
        
            \filldraw[fill=midgrey!20] (-0.2, 1.5) rectangle (10.2,0.5);
    \node (initialLte) at (5,1) { Scheduled Communication };

  \end{tikzpicture}
  \caption{Proposed uplink multi-connectivity initial access procedure. Red and green dashed lines refer to the control messages exchanged via the bachaul X2 and the legacy communication links, respectively.}
  \label{fig:initial_access_procedure_fig}
\end{figure}
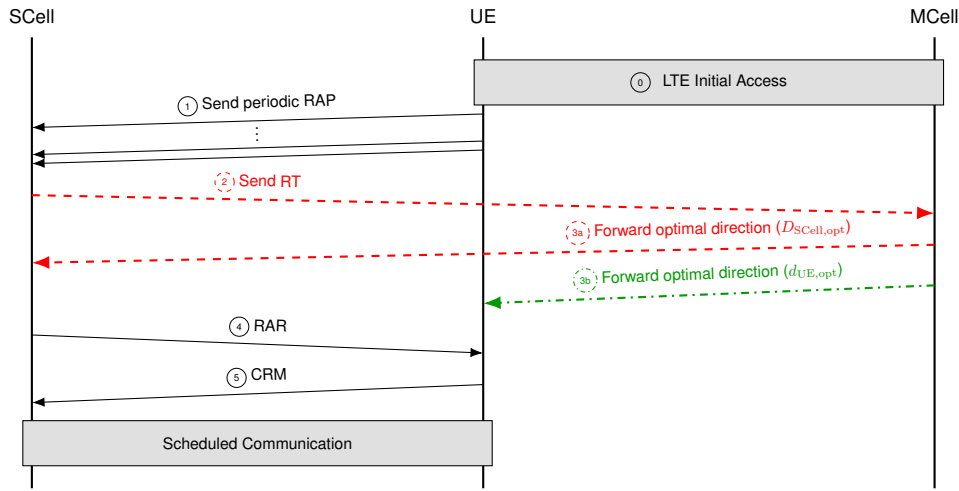

\begin{enumerate}
\item[0)] A user  searches for
synchronization signals from conventional 4G cells.  This detection is~fast since it
can be performed omnidirectionally and there is no need for directional~scanning \footnote{
Under the assumption that the
5G mmWave eNBs are roughly time synchronized to the 4G cell, and the round trip propagation times
are not large, an uplink transmission from the UE will be roughly time aligned at any
closeby mmWave cell.  For example, if the cell radius is 150~m (a typical mmWave cell),
the round trip delay is only 3~$\mu$s.}.

\item A user desiring initial access
 broadcasts a \emph{random access preamble} (RAP) scanning different angular directions, while the mmWave cells scan for the presence  of those messages.
  According to the MC procedure's first phase, multiple RTs are collected at the SCell sides.
Each of these RAPs will arrive roughly time-aligned in the random access slots of
all potential neighboring mmWave cells.

\item Each SCell forwards its RT to the MCell via the X2 backhaul link.
 In analogy with the MC's second phase, the MCell  performs the best attachment decision, based on the received RTs, together with selecting the optimal directions for the transceiver to communicate.

 \item The MCell forwards to the designated SCell (via the usual high capacity backhaul) and to the user (via the legacy LTE band though an omnidirecional control message) the respective sectors through which they should steer the beam to communicate.

\item At this point, the best SCell-UE beam pair has been determined, therefore both the user and the SCell can steer through their optimal directional sectors, obtaining the full beamforming gain.
So, the SCell transmits a \emph{random access response} (RAR) to the UE, containing some initial timing and power correction information.

\item After receiving the RAR, the UE sends a \emph{connection request message} (CRM) on the resources scheduled in the  grant in the RAR.
All subsequent communication can occur on scheduled channels. As in 3GPP LTE, the immediate subsequent messages would be used for connection set up and contention resolution \cite{Barati}.

\end{enumerate}

The performance of the presented IA procedure will be analyzed in detail in Sec. \ref{sec:IA_res}. Anyway, we immediately notice that  the attachment decision is made by the MCell, which oversees the whole network and collects channel reports from all the surrounding mmWave cells.
Therefore, unlike in  traditional attachment policies, the association can be possibly performed by accounting for the instantaneous load conditions of the neighboring cells, to guarantee  enough fairness and reliability to the whole cellular network.
Furthermore, we claim that an uplink-based scheme allows for faster IA than its downlink counterpart, especially when a beamforming architecture with multiple analog streams is preferred for the sweeping operations.

%
%

\subsection{RLF Detection and Recovery}
\label{sec:RLF}

One of the key challenges  systems operating in  mmWave bands have to cope with is the rapid channel dynamics. Unlike in conventional LTE systems, mmWave signals are completely blocked by many common building materials such as brick and mortar. As a result, the movement of obstacles and reflectors, or even changes in the orientation of a handset relative to a body or hand, can cause the channel to rapidly appear or disappear \cite{rappaport2014millimeter}. When a radio-link failure (RLF) occurs, the link that has been established between user and eNB is obstructed, with a consequent  SINR and throughput degradation. The UE should immediately react by adapting its beam pair or, as a last resort, by triggering a handover~\cite{EW2016}.

Most literature refers to challenges that have been recently analyzed in the 60 GHz IEEE 802.11ad WLAN and WPAN scenarios. In \cite{block_60}, for example, a detailed investigation of the  effect of people movement on the temporal fading envelope is performed. Some related works present different solutions to address the blockage issue described above, such as \cite{steering_closed,smart_BF,ferrante2015mm}.

\begin{figure}[t!]
\centering
 \includegraphics[trim= 0cm 0cm 0cm 0cm , clip=true, width=0.75 \textwidth]{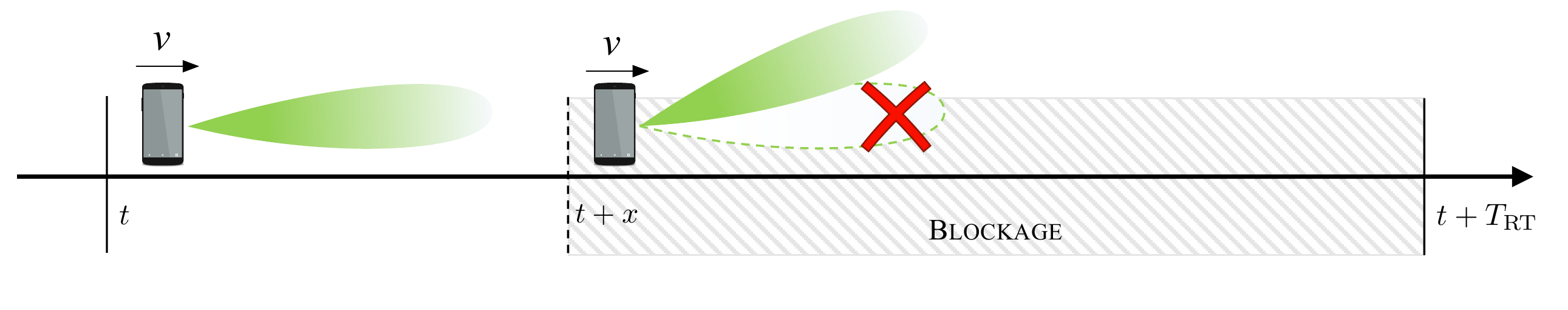}
 \caption{MC procedure for RLF recovery. At time $t$ and $t+T_{\rm RT}$ the SCell collects a RT. At time $t+x$ a blockage event occurs and the user, moving at constant velocity $v$, loses the connection with its current serving mmWave SCell. The UE and the SCell can promptly react to the channel failure by exploiting previously saved report table entries.}
 \label{fig:block}
\end{figure}

In the event that the primary propagation path is obstructed, the MC procedure presented in Sec. \ref{sec:MCP} can be employed to partially overcome the link failure leveraging on: (i) the presence of the central controller and the very  reliable and stable  LTE connection the UEs can benefit from, and (ii) previously saved instances of the channel quality information exchanged by the network nodes.
 We use Fig. \ref{fig:block} as an example.
  Assume that, at time $t$, the user moving at constant speed $v$  is connected to a specific mmWave SCell, through direction $d_{\rm UE,opt}$.
 We assume that, at time $t + x$ and before a new RT is generated, a blockage occurs.
If no practical actions are taken, the user has to wait for a new RT to be collected (at time $t+T_{\rm RT}$) before a new optimal beam pair, able to circumvent the obstruction, is determined. 
Indeed, during $T_{\rm RT} - x$ s, the user's experienced rate is zero, due to the link breakdown and the consequent SINR collapse.
One practical solution is to immediately react to the path impairment by taking advantage of previously saved instances of the RT.
As soon as a blockage is detected, the UE can autonomously access its  most recent RT entries (or set of previous tables) and find the second best direction $d_{\rm UE, subopt}  \neq d_{\rm UE, opt}$ to communicate, as a sort of \emph{backup procedure} before the transceiver fully recovers the optimal beam configuration. Such beam pair will be a suboptimal solution (since the optimal path is blocked), but at least allows the user to experience a higher average throughput than  it would have achieved if no actions were taken. 

Having a second available link, when the primary path is obstructed,  adds  diversity and robustness to the communication.
In Sec. \ref{sec:RLF_res}, we will numerically show the advantages, in terms of throughput, of establishing a backup beam configuration between UE and eNB, after a RLF is detected, rather than just waiting for a complete network decision to be made.

\section{Simulation Parameters Description}
\label{sec:param_descr}
In Sec. \ref{sec:channel_model}, we describe the mmWave and the LTE channel models we used to run the simulations, and in Sec. \ref{sec:mobility_model} we present the mobility model we employed to account for the dynamics which affect the mmWave propagation. Finally, in Sec. \ref{sec:sim_param}, we present our main simulation parameters.

\subsection{Channel Models}
\label{sec:channel_model}

\textbf{Mmwave Channel Model.} The channel model we have implemented is based on recent real-world measurements at $28$ GHz in New York City, to provide a realistic assessment of mmWave micro and picocellular networks in a dense urban deployment.
The parameters  that are used to generate one instance of
the channel matrix \textbf{H} include: (i) spatial clusters; (ii) fractions
of power; (iii) angular beamspreads; and (iv) a small-scale
fading model, massively affected by the Doppler shift, where
each of the path clusters is synthesized with a large number
of subpaths. A complete description of the channel parameters
can be found in \cite{Mustafa,MacCartney2015Wideband,ns3_nokia}.

The distance-based pathloss, which models Line-of-Sight (LoS), Non-Line-of-Sight (NLoS) and outage, is defined as
$PL(d)[dB] = \alpha + \beta 10 \log_{10}(d)$,
where $d$ is the distance between the receiver and the transmitter and the values of the parameters $\alpha$ and $\beta$ are given in \cite{Mustafa}.

Due to the high pathloss experienced at mmWaves, multiple antenna elements with beamforming (BF) are essential to provide an acceptable  communication range. The BF gain from transmitter $i$ to receiver $j$ is  given by:
\begin{equation}
G_{\rm BF}({{i,j}}) = |\textbf{w}^*_{rx_{i,j}}\textbf{H}_{ij}\textbf{w}_{tx_{i,j}}|^2
\label{beamforming_gain}
\end{equation}
where $\textbf{H}_{i,j}$ is the channel matrix of the $ij^{th}$ link, $\textbf{w}_{tx_{i,j}}\in \mathbb{C}^{n_{\mathbb{T}x}}$ is the BF vector of transmitter $i$ when transmitting to receiver $j$, and $\textbf{w}_{rx_{,ij}}\in \mathbb{C}^{n_{\mathbb{R}x}}$ is the BF vector of receiver $j$ when receiving from transmitter $i$.  
Analog or digital beamforming architectures are typically considered. 
The former shapes the output beam with only one radio frequency (RF) chain, using
phase shifters. This model saves power by using only a single Analog-to-Digital Converter (ADC)
but has limited flexibility since the eNBs can only beamform in one direction at a time.
On the other hand, the latter configuration provides the highest flexibility in shaping the 
beams,  allowing transmission/reception in multiple directions simultaneously. However, it requires one RF chain per antenna element, thus increasing the cost and complexity of the architecture  \cite{sun2014mimo}.

The channel quality is measured in terms of SINR. By referring to the mmWave statistical channel described above, the SINR between eNB$_m$ and a test UE is:
\begin{equation}
\text{SINR}( m) = \frac{\frac{P_{\rm TX_{mmW}}}{PL_{ m}}G_{\rm BF}{(m, \rm UE)}}{\sum_{k\neq m}\frac{P_{\rm TX_{mmW}}}{PL_{ k}}G({ k,\rm UE})+W_{\rm mmW}\times N_0}
\vspace{0.5cm}
\label{eq:SINR}
\end{equation}
where $G_{\rm BF}({ m,\rm UE})$ and $PL_{ m}$ are the BF gain and the pathloss obtained between eNB$_{m}$ and the UE, respectively, and ${ W_{\rm mmW}\times N_0}$ is the thermal noise.
In \eqref{eq:SINR}, it is assumed that the UE is interfered by other transmitters. However, to some extent, given the wide bandwidth, it is easy to orthogonalize the SRSs across multiple users and we can assume that the SRS waveforms are transmitted over multiple sub-signals with each sub-signal being transmitted over a small bandwidth $W_{\rm sig}$. 
The use of the sub-signals can provide frequency diversity, and narrowband signals in~the control plane  remove any inter-cell interference and support  low power receivers with high SINR capabilities~\cite{Barati}.

Finally, the rate ($R$) experienced by the UE connected to eNB$_{m}$ is approximated using the Shannon capacity:
\begin{equation}
R({m)} = \frac{W_{\rm mmW}}{N_m}  \log_2\Big(1+\text{SINR}(m)\Big)
\label{eq:rate}
\end{equation}

where $N_m$ is the  number of users that are currently being served by eNB$_m$ and $W_{\rm mmW}$ is the available total bandwidth.

\textbf{LTE Channel.} A connection to the LTE band is required when the mmWave primary propagation path is obstructed or not available, or to reliably forward the scheduling/attachment decisions to the final user.
According to the LTE 3GPP specifications in \cite{3GPP_DC} and considering an outdoor dense scenario, a distance-depended pathloss from the MCell to the UE can be defined.
In particular, assuming that the user, being at distance $R$ from the LTE eNB, is in a LoS pathloss state, with probability
\beq
\mathbb{P}_{\rm LoS}(R) = \min \Big( \tfrac{0.018}{R},1 \Big) \left[ 1-\exp\left(\frac{-R}{0.063}\right)\right]+
\exp\left(\frac{-R}{0.063}\right),
\eeq
the LoS pathloss can be defined as:
\beq
PL_{\rm LoS}(R) = 103.4+24.2\log_{10}(R).
\eeq
Conversely, if a NLoS condition holds (with probability $\mathbb{P}_{\rm NLoS}(R)=1-\mathbb{P}_{\rm NLoS}(R)$), the NLoS pathloss can be defined as:
\beq
PL_{\rm NLoS}(R) = 131.1+42.8\log_{10}(R).
\eeq
When considering an LTE connection, signals are assumed to be exchanged through omnidirectional channels. Therefore, if we deploy just one LTE eNB in the reference scenario, the quality of the received information is measured, in terms of SNR,~by:
\begin{equation}
\text{SNR} = \frac{P_{\rm TX_{LTE}}/PL(R)}{W_{\rm LTE}\times N_0},
\vspace{0.5cm}
\label{eq:SNR}
\end{equation} 
where $PL(R)$ is the pathloss (either LoS or NLoS) experienced between the MCell and the test UE, and $W_{\rm LTE}\times N_0$ is the thermal noise. The rate can be computed according to Eq. \eqref{eq:rate}.

\subsection{Mobility Model}
\label{sec:mobility_model}
One of the key challenges for cellular systems in the mmWave bands is the rapid channel dynamics. When moving, the user experiences a strong Doppler shift  whose effect increases with speed.
However, most channel studies have been performed in stationary locations with minimal local blockage, making it difficult to estimate the fluctuations that affect a realistic mmWave environment.
In order to simulate such dynamics, we propose a mobility model in which the small and the large scale fading parameters of the mmWave \textbf{H} matrix  are periodically updated, to emulate short variations and sudden changes of the perceived channel, respectively.

The Doppler shift and the spatial signatures are updated at every time slot, according to the user speed and its position, in terms of angle of arrival (AoA) and departure (AoD). The distance-based pathloss is also updated, but we maintain the same pathloss state (LoS, NLoS or outage) recorded in the previous complete update of the \textbf{H} matrix.
On the other hand,  to capture the effects of the long term fading, the \textbf{H} matrix parameters (i.e., the number of spatial clusters and subpaths, the fractions of power, the angular beamspreads and the pathloss conditions) are completely updated every  $T_H$ s, for all the mmWave links between each UE and each SCell.
 We recall that this may cause the user to switch from a certain pathloss state to another  (e.g., from LoS to NLoS, to simulate the presence of an obstacle between transmitter and receiver), with a consequent sudden drop of the channel quality by many dBs.


 The beamforming vectors are not adapted when the \textbf{H} matrix is updated. We need to wait for a new  RT to be collected (every $T_{\rm RT}$ s) to detect the (possibly changed) channel propagation conditions and properly react, i.e.,   by adapting the directions through which the UE and the designated SCell steer their beams.
 Frequent RTs (small $T_{\rm RT}$) and flat channels (large $T_H$) result in a good monitoring of the user and good average channel gains.
 In Sec. \ref{sec:HO_res} we  show how  variations of $T_{\rm RT}$ and $T_H$ affect  the communication quality.
%
%

\begin{figure}[t!]
	\centering
		\setlength{\belowcaptionskip}{0cm}
		\setlength{\belowcaptionskip}{0cm}
		\setlength\fwidth{0.9\textwidth}
		\setlength\fheight{0.2\textwidth}
		\input{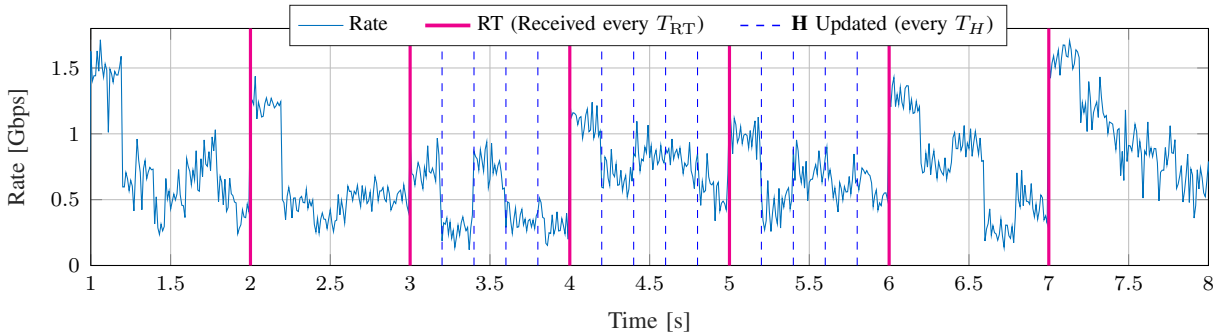}	
		\caption{Example of time-varying rate experienced by a user moving at  speed $v=20$ m/s in a mmWave scenario in which ${M=70}$~BS/km$^2$ are deployed. The RTs are generated every $T_{\rm RT} = 1$ s (vertical magenta lines), the small scale fading parameters of the channel  vary every time slot of $1$ ms, the large scale fading parameters of the channel matrix \textbf{H} vary every $T_H = 200$ ms (vertical blue dotted lines).}
		\vspace{-1.2em}
		\label{fig:example_rate}
\end{figure}
 
As an example, in Fig. \ref{fig:example_rate}  we plot the rate experienced by a test user,  moving at speed $v=20$ m/s  along a straight line during an excerpt of  a simulation. 
The large scale fading parameter of the mmWave channel are updated every $T_H=200$~ms, while the communication configuration may be updated every $T_{RT}=1$ s, upon the dissemination of the RTs to the MCell.
For instance, it can be seen that, at time $t=2^-$, the rate has strongly degraded,  since the user has moved without updating its beam steering direction and thus has misaligned from its serving SCell. 
However, at time $t=2$, a new CRT has been generated and the transceiver is finally able to update its beam configuration (by performing a beam switch operation) or the user can handover (by choosing a serving mmWave eNB providing better communication performance), thus recovering the maximum  achievable transmission rate.
We notice that wide rate collapses (e.g., at time $t=3.2$ or $t=5.2$) mainly refer to pathloss state changes (i.e., from LoS to NLoS), caused by the update of the large scale fading parameters of \textbf{H}, while the rapid fluctuations of the rate are due to the adaptation of the small scale fading parameters of the channel (and mainly to the Doppler effect experienced by the moving user).

\subsection{Simulation Scenario}
\label{sec:sim_param}

The parameters used to run our simulations are based on realistic system design considerations and are summarized in Tab. \ref{tab:params}.
\renewcommand{\arraystretch}{0.6}
\begin{table}[!t]
\small
\centering
\caption{Main simulation parameters.}
\begin{tabularx}{0.8\textwidth}{ @{\extracolsep{\fill}} lll}
\toprule
Parameter & Value & Description \\ \midrule
 $W_{\rm mmW}$ & $1$ GHz & Bandwidth of mmWave eNBs\\
 
$f_{\rm c, mmW}$ & $28$ GHz &  mmWave carrier frequency \\

$P_{\rm TX, mmW}$ & $30$ dBm & Transmission power of mmWave eNBs \\

 $W_{\rm LTE}$ & $20$ MHz & Bandwidth of LTE eNB\\
 
$f_{\rm c, LTE}$ & $2$ GHz &  LTE carrier frequency \\

$P_{\rm TX, LTE}$ & $46$ dBm & Transmission power of LTE eNB \\


$\Gamma_{\rm out}$ & $ -5$ dB &  Minimum SINR threshold \\

$N_{\rm ANT,SCell}$ & $8 \times 8$  & SCell UPA MIMO array size  \\

$N_{\rm ANT,UE}$ & $4 \times 4$ & UE UPA MIMO array size\\

$N_{\rm SCell}$& $16$  & SCell scanning directions  \\

$N_{\rm UE}$& $8$  & UE scanning directions  \\

$T_{\rm sim}$ & $10$ s & Simulation duration \\

$v$ & $20$ m/s & UE speed\\

$M$ & Varied & mmWave SCell density \\

$N_m$ & 10 & Users per mmWave SCell\\

$T_{\rm sig}$ & $10 \, \: \mu s$& SRS duration \\

$\phi_{\rm ov}$ & $5\%$ & Overhead\\

$T_{\rm per}$ & $200 \: \mu$s & Period between PSS transmissions \\

$T_{H}$ & Varied & Channel update periodicity\\

$T_{RT}$ & Varied & Time between two consecutive RTs\\

$T_{B}$ & Varied & Blockage duration\\
\bottomrule
\end{tabularx}
\label{tab:params}
\end{table}
Our results are derived through a Monte Carlo approach, where multiple independent simulations of duration $T_{\rm sim}$  are repeated, to  get different statistical quantities of interest. In each experiment: (i) under the control of one single MCell operating in  the legacy band, we deploy $M$ mmWave SCells and $N$ UEs, according to a Poisson Point Process (PPP) and as done in \cite{Heath}, with an average density of $N_m=10$ users per cell (as foreseen in \cite{3GPP_5G} for a dense urban environment); (ii) we run the multi-connectivity framework described in  Sec. \ref{sec:MCP}  by establishing a mmWave link between each SCell-UE pair and collecting the SINR values at each SCell, according to Eq.~\eqref{eq:SINR}, when the transceiver performs the sequential scan; and (iii) we select the most profitable mmWave eNB the user should attach to, according to either a maximum SINR or maximum rate policy.

We  consider an SINR threshold $\Gamma_{\rm out} = -5$ dB, assuming that, if ${\text{SINR}_{i,j}(m) < \Gamma_{\rm out}}$, no control signals are collected when the UE  transmits through direction $i$  and the BS $m$ is receiving through direction $j$.
Decreasing $\Gamma_{\rm out}$ would allow finding more SCells, at the cost of designing more complex (and expensive) receiving schemes, able to detect the intended signal in more noisy channels. If a multi-connectivity approach is chosen, the UE can still reach the MCell (by establishing a connection over the LTE band) when the signal quality is below $\Gamma_{\rm out}$.

 A set of two dimensional antenna arrays is used at both the mmWave eNBs and the UE.
SCells are equipped with a Uniform Planar Array (UPA) of $8 \times 8$ elements, which allow them to steer beams in $N_{\rm SCell}=16$ directions; the user exploits an array of $4 \times 4$ antennas, steering beams through $N_{\rm UE}=8$ angular directions\footnote{In this work, we assume a 2D structure for the cells. 
Nevertheless, our system is easily customizable and allows for the design of an advanced 3D scanning technique as well.
 However, such a choice would  lead to an  increase of the time required to complete each iteration of the presented measurement reporting scheme, without providing any further noticeable~insights.}. 
The spacing of the elements is set to $\lambda/2$, where $\lambda$ is the wavelength.

In the first phase of the presented sweeping algorithm, we alternate portions of time in which SRSs are periodically transmitted in  brief intervals of length $T_{\rm sig}$, and intervals of length ${T_{\rm per} \gg T_{\rm sig}}$ in which each eNB and each UE handle their usual traffic operations. 
We took $T_{\rm sig} = 10$ $\mu s$, which is sufficiently small that the channel will be coherent even at the very high frequencies for mmWaves, and $T_{\rm per} = 200$ $\mu s$, in order to maintain a constant overhead of $\phi_{\rm ov}=5\%$.  In this way, the mmWave eNBs will be taken out of their standard communication capabilities  only for $\phi_{\rm ov}=T_{\rm per} / T_{\rm sig} =5 $ percent of their operational~time\footnote{The values of $T_{\rm per}$ and $T_{\rm sig}$ have been chosen according to the analysis in \cite{Barati,CISS,magazine_IA}.  
However, the proposed framework is general and its parameters can be tuned according to the  peculiarities of any specific simulation environment.}.

\section{Results and Performance Evaluation}
\label{sec:res}

In this section, we present some simulation results aiming at:
\begin{itemize}
\item[(i)] comparing the performance of the uplink multi-connectivity framework described in Sec. \ref{sec:MCP} with a traditional downlink standalone approach in terms of delay, data rate,  energy consumption and stability;
\item[(ii)] giving numerical evidence of the performance of several control applications (i.e., handover, beam tracking, initial access, RLF recovery) which can be enabled in next-generation  mmWave systems by the presented measurement reporting framework;
\item[(iii)] showing how the variability of the mmWave channel affects the performance of a cellular network (mainly in terms of achievable throughput).
\end{itemize}

\subsection{Comparison with Downlink Standalone Scheme}
\label{sec:comparison_DL}

\textbf{Delay.}
We define as $D$ the time \emph{delay} required to complete each iteration of 
either the uplink multi-connectivity framework presented in this work or a traditional downlink standalone scheme.
We claim that the first phase of the proposed framework (i.e., uplink measurements) dominates the overall delay performance, provided that: 
(i) the switching time for beam switching is in the scale of nanoseconds, and so it can be neglected~\cite{chandra2014adaptive}; (ii) in the second phase of the procedure, according to \cite{JSAC_2017},  RTs are sent through the X2 links, which may be wired or wireless backhaul and whose latency is assumed to be negligible. In addition, the backhaul overhead is minimal since it requires only transmission of Channel QuaIity Indicator (CQI) levels for each cell, which is very small compared to the data;
(iii) in the third phase of the procedure, the attachment decisions are once again  forwarded to the SCells through the high capacity backhaul and through  omnidirectional LTE messages, whose latency is ignored if the UEs  have already set up a link to the MCell.

Following \cite{Barati,barati2015initial},
we suppose that in either the uplink or the downlink
direction, the synchronization signals are $T_{\rm sig}$ long
and occur once ever $T_{\rm per}$ s.
  The size of $T_{\rm sig}$ is
determined by the necessary link budget and we will assume that it is the same in both directions.  The values in Tab.~\ref{tab:params} are based on simulations in
\cite{barati2015initial} that enable reliable detection with an overhead of $T_{\rm sig}/T_{\rm per}$
of 5\%.  Now, the scanning for the synchronization
signal for each SCell-UE direction will require
$N_{\rm SCell}N_{\rm UE}/L$ scans, where $L$ is the number of directions in which the receiver can look at any one time.  Since there is one scanning opportunity every $T_{\rm per}$
s, the total delay is
\beq
	D = \frac{N_{\rm SCell}N_{\rm UE}T_{\rm per}}{L}.
	\label{eq:D}
\eeq

\begin{table}[t!]
\centering
\renewcommand{\arraystretch}{1.2}
\caption{Delay $D$ to complete each iteration of either the uplink multi-connectivity measurement framework described in Sec.~\ref{sec:MCP} or a traditional downlink standalone approach.  A comparison among different BF architectures (analog and fully digital) is performed. We assume $T_{\rm sig} = 10 \: \mu$s, $T_{\rm per} = 200 \: \mu$s (to maintain an overhead $\phi_{\rm ov} = 5\%$), $N_{\rm SCell} = 16$ and~$N_{\rm UE} = 8$.}
\begin{tabular}{c |c| c}
\hline
\makecell{\textbf{BF Architecture}} & \makecell{\textbf{UL Scheme} \\  \footnotesize UEs transmit \\[0.01\baselineskip]  \footnotesize SCells receive} & \makecell{\textbf{DL Scheme}\\ \footnotesize UEs receive \\ \footnotesize SCells transmit}\\
 \hline  \hline
 Analog& \cellcolor{blue!0}$N_{\rm SCell} N_{\rm UE} T_{\rm per} $ ($25.6$ ms) & $N_{\rm SCell} N_{\rm UE} T_{\rm per} $ ($25.6$ ms)  \\
  \hline
\makecell{Digital$^{(*)}$}& \cellcolor{blue!10} $\frac{N_{\rm SCell} N_{\rm UE} T_{\rm per}}{N_{\rm SCell}} $ ($1.6$ ms)  &  $\frac{N_{\rm SCell} N_{\rm UE} T_{\rm per}}{N_{\rm UE}} $  ($3.2$ ms)\\
  \hline
\end{tabular}
\begin{tablenotes}
      \scriptsize
      \item $^{(*)}$ Digital beamforming is applied at the receiver  (i.e., at the SCell side if an UL scheme is preferred, or at the UE side otherwise.)
    \end{tablenotes}
\label{tab:D}
\end{table}

The value of $L$ depends on the beamforming capabilities (and the array size).  In the uplink-based design, $L=1$ if
the SCell receiver has analog BF and $L=N_{\rm SCell}$ if it has a fully digital transceiver.
Similarly, in the downlink,
$L=1$ if the UE receiver has analog BF and $L=N_{\rm UE}$ if it has a fully digital transceiver.
Tab. \ref{tab:D} compares the resulting delays for UL- and DL-based designs
depending on the BF capabilities of the receiver.
We see that the UL design offers a significantly reduced access delay when a digital architecture is preferred and makes it possible to complete every repetition of the measurement reporting framework every at least $1.6$ ms (when considering an overhead $\phi_{\rm ov}=5\%$).
The main reason is that we usually consider $N_{\rm SCell} \gg N_{\rm UE}$, due to the base station's less demanding space constraints with respect to a mobile terminal: a larger number of antenna elements can be packed at the eNB side, with a consequently number of directions that can potentially be scanned simultaneously through a digital beamforming scheme.

\textbf{Rate.}
In Tab. \ref{tab:R}, we evaluate the average rate $\mathbb{E}[R]$ experienced by a test user when either a multi-connectivity or a traditional standalone mobility management framework is applied, for different SCell density values.
It can be observed that the rate achievable
with the first solution is  higher than with the second one.
 The reason is that, when relying on the LTE
eNB for dealing with outage events, the UE experiences a non-zero
throughput, in contrast to the standalone configuration
which cannot properly react to a situation where no mmWave
SCells are within reach. 
The gap between the two architectures is quite remarkable when considering very sparse environments (i.e., $M<20$ SCell/km$^2$).
In those scenarios, most users would be in an outage pathloss status, making the fallback to the legacy connectivity a vital option to recover a  sustainable communication quality. However,  achievable rates at those densities are very low due to the quite reduced data rate that a low-bandwidth, overloaded LTE~eNB can offer.

We notice that $\mathbb{E}[R]$ increases with the SCell density $M$. In fact, the inter-cell distance is reduced and each UE generally finds a closer SCell (showing better channel propagation conditions) to associate with, thus experiencing an increased SINR (and rate) too.
No rate difference is registered between the multi-connectivity and the standalone configurations when considering very dense environments. In those circumstances, users will not suffer an outage and their traffic will be properly handled by the available mmWave SCells, without necessarily having to fallback to the LTE eNB.

\begin{table}[t!]
\centering
\begin{minipage}{.45\textwidth}
\centering
\renewcommand{\arraystretch}{1.2}
\captionaboveof[ee]{table}{Rate $\mathbb{E}[R]$ experienced when either the multi-connectivity framework described in Sec.~\ref{sec:MCP} or a   standalone approach is used. $T_{H} = 100$~ms, ${T_{RT} = 300}$~ms. }
\begin{tabular}{c |c| c}
\hline
\makecell{$M$  [SCell/km$^2$]} & \makecell{\textbf{Multi-connectivity}} & \makecell{\textbf{Standalone}}\\
 \hline  \hline
 4&\cellcolor{blue!25} $8.19$ Mbps & $5.5$ Mbps  \\
  \hline
10& \cellcolor{blue!20}$41.09$ Mbps & $35.4$ Mbps  \\
    \hline
    20& \cellcolor{blue!15}$122.9$ Mbps & $121$ Mbps  \\
    \hline
    40&\cellcolor{blue!10} $360.42$ Mbps & $357.6$ Mbps  \\
    \hline
    70& \cellcolor{blue!5}$778.87$ Mbps & $777.4$ Mbps  \\
    \hline
    90& \cellcolor{blue!0}$1103.6$ Mbps & $1103.6$ Mbps  \\
    \hline
\end{tabular}
\label{tab:R}
\end{minipage}\qquad \qquad 
\begin{minipage}{.45\textwidth}
\captionaboveof[ee]{table}{  $E_C$ to complete each iteration of either the uplink multi-connectivity measurement framework described in Sec.~\ref{sec:MCP} or a traditional downlink standalone approach. A digital BF configuration is~applied at the receiver side. $T_{\rm per} = 200 \: \mu$s, $N_{\rm SCell} = 16$ and~$N_{\rm UE} = 8$. }
\centering
\renewcommand{\arraystretch}{1.2}
\begin{tabular}{c |c| c}
\hline
\makecell{\textbf{Network} \\ \textbf{Entity}} & \makecell{\textbf{UL Scheme} \\  \footnotesize UEs transmit \\[0.01\baselineskip]  \footnotesize SCells receive} & \makecell{\textbf{DL Scheme}\\ \footnotesize UEs receive \\ \footnotesize SCells transmit}\\
 \hline  \hline
 SCell& $1.7477$ J & $ 0.0665$ J  \\
  \hline
UE& \cellcolor{blue!10} $0.0287$ J &$0.8739$ J\\
    \hline
\end{tabular}
\label{tab:EC}
\end{minipage}
\end{table}

Finally, rate gains will likely be even more significant for increasing values of $T_{\rm RT}$. In fact, less frequent tracking operations might lead to a more remarkable channel degradation between the transmitter and the receiver, making the fallback to LTE an increasingly attractive option to restore an adequate communication quality.

\textbf{Energy Consumption.}
The energy consumption ($E_C$) of each mobility management configuration can be evaluated as the product between the  power ($P_C$) and the time delay ($D$) required to complete each iteration of each approach
\footnote{The total power consumption ($P_C$) of each beamforming scheme is evaluated according to  \cite{Waqas_EW2016, abbas2016_ECIA}, in which  $b = 3$ quantization bits are used by the Analog-to-Digital Converter block.}.
According to Tab.~\ref{tab:D}, when considering an uplink multi-connectivity scheme,   digital BF is used at the SCell side while  analog BF is preferred at the UE side, and $D^{\rm MC}=1.6$ ms. Therefore:
\begin{equation}
E_{C, \rm SCell}^{\rm MC} = P_{C}^{\rm DBF} \cdot D^{\rm MC} \qquad  \qquad E_{C, \rm UE}^{\rm MC} = P_{C}^{\rm ABF} \cdot D^{\rm MC}
\end{equation}
For a downlink standalone configuration, analog BF is used at the SCell side while  digital BF is preferred at the UE side, and $D^{\rm SA}=3.2$ ms. Therefore
\begin{equation}
E_{C, \rm SCell}^{\rm SA} = P_{C}^{\rm ABF} \cdot D^{\rm SA} \qquad  \qquad E_{C, \rm UE}^{\rm SA} = P_{C}^{\rm DBF} \cdot D^{\rm SA}
\end{equation}
In Tab. \ref{tab:EC}, we compare the energy performance of the two approaches. 
It is evident that, although the UL scheme is more consuming at the SCell side, it is more energy efficient at the UE side. 
This represents a very relevant feature of the proposed multi-connectivity framework since mobile terminals are the most \emph{energy-constrained} network entities, due to their limited battery capacity (contrary to the infrastructure nodes which are always power connected and do not suffer from strict energy requirements). 
We therefore claim that an UL framework, able to reduce the energy consumption of the mobile terminal by around 30 times (with the settings of Tab. \ref{tab:EC}) with respect to its DL counterpart, should therefore be preferred to enable a more efficient mobility management scheme.

\textbf{Robustness.}
In order to compare the robustness of the multi-connectivity and the standalone configurations, following the analysis we proposed in \cite{JSAC_2017}, we use the~ratio 
\beq
R_{\rm var} = \frac{\sigma_{R}}{\mathbb{E}[ R]},
\label{eq:var}
\eeq
where $\mathbb{E}[R]$ is the mean value of the  throughput measured for each approach and  $\sigma_{R}$ is its standard deviation.
High values of $R_{\rm var} $ reflect remarkable channel instability, thus the rate would be affected by local variations and periodic degradations.

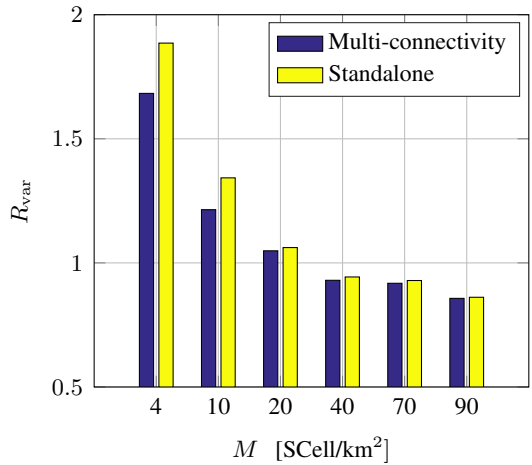
\begin{figure}[t!]
     \centering
     		\setlength{\belowcaptionskip}{0cm}
	\setlength{\belowcaptionskip}{0cm}
	\setlength\fwidth{0.35\textwidth}
	\setlength\fheight{0.3\textwidth}
%
%
\usetikzlibrary{spy}
\definecolor{mycolor1}{rgb}{0.20810,0.16630,0.52920}%
\definecolor{mycolor2}{rgb}{0.97630,0.98310,0.05380}%

\tikzstyle{every pin}=[fill=white,
draw=black,
font=\footnotesize]

\pgfplotsset{
tick label style={font=\footnotesize},
label style={font=\footnotesize},
legend  style={font=\footnotesize}
}

\begin{tikzpicture}

\begin{axis}[%
width=\fwidth,
height=\fheight,
at={(0\fwidth,0\fheight)},
scale only axis,
bar shift auto,
xmin=0,
xmax=7,
xtick={1,2,3,4,5,6},
xticklabels={{4},{10},{20},{40},{70},{90}},
xlabel style={font=\color{white!15!black}},
xlabel={$M$ $ \text{ [SCell/km}^2]$},
ylabel style={font=\color{white!15!black}},
xlabel style={font=\color{white!15!black}},
ymin=0.5,
ymax=2,
ylabel={$R_{\rm var}$},
label style={font=\footnotesize},
axis background/.style={fill=white},
xmajorgrids,
ymajorgrids,
legend style={legend cell align=left, align=left, at={(0.98,0.98)}, draw=white!15!black, legend columns = 1}]

\addplot[ybar, bar width=0.229, fill=mycolor1, draw=black, area legend] table[row sep=crcr] {%
1 1.6830   \\  
2 1.2144   \\ 
3 1.0487    \\
4 0.9299    \\
5 0.9179\\
6  0.8571\\
};
\addlegendentry{Multi-connectivity}

\addplot[ybar, bar width=0.229, fill=mycolor2, draw=black, area legend] table[row sep=crcr] {%
1 1.8856    \\ 
2 1.3426   \\ 
3 1.0617   \\ 
4 0.9433    \\
5 0.9290\\
6  0.8615\\
};
\addlegendentry{Standalone}


\end{axis}

%
\end{tikzpicture}%
        \caption{\footnotesize Average ratio $R_{\rm var}$ vs. SCell density, showing the stability of the channel during the simulation.}
        \label{fig:var}
 \end{figure}

Let $R_{\rm var}^{\rm MC}$ and $R_{\rm var}^{\rm SA}$ be the variance ratios of Eq. \eqref{eq:var} for the multi-connectivity and the standalone configurations, respectively. 
From Fig. \ref{fig:var}, we observe that $R_{\rm var}^{\rm MC}$ is lower than $R_{\rm var}^{\rm SA}$, for each value of  density $M$, making it clear that the LTE eNB employed in a MC configuration can stabilize the rate, which is not subject to significant variations. 
In fact, in the portion of time in which the UE would experience zero gain if a standalone architecture were implemented (due to an outage event), the rate would suffer a noticeable discrepancy with respect to the LoS values, thus increasing the rate variance throughout the simulation. This is not the case for the MC configuration, in which the UE can always be supported by the LTE eNB, even when a blockage event affects the scenario.
This result is fundamental for real-time applications, which require a long-term stable throughput to support high data rates and a consistently acceptable Quality of Experience for the users.

Finally, we observe that, in general, the stability of the network rate increases with $M$ (showing smaller values of $R_{\rm var} $), due to the more comparable  values of SINR (and rate) that are guaranteed.
Furthermore, in denser environments and as the probability of pathloss outage decreases, the gap between the two configurations decreases, as the role of the LTE eNB becomes less relevant.

\subsection{Handover Performance}
\label{sec:HO_res}

\begin{figure}[t!]
     \centering
     		\setlength{\belowcaptionskip}{0cm}
	\setlength{\belowcaptionskip}{0cm}
	\setlength\fwidth{0.95\textwidth}
	\setlength\fheight{0.34\textwidth}
%
%
\definecolor{mycolor1}{rgb}{0.20810,0.16630,0.52920}%
\definecolor{mycolor2}{rgb}{0.07935,0.52002,0.83118}%
\definecolor{mycolor3}{rgb}{0.21783,0.72504,0.61926}%
\definecolor{mycolor4}{rgb}{0.85066,0.72990,0.33603}%
\definecolor{mycolor5}{rgb}{0.97630,0.98310,0.05380}%
\usetikzlibrary{spy}

\begin{tikzpicture}

\pgfplotsset{
tick label style={font=\footnotesize},
label style={font=\footnotesize},
legend  style={font=\footnotesize}
}

\begin{axis}[
xtick={1,2,3,4,5},
xticklabels={{0.01},{0.1},{0.2},{0.5},{1}},
width=\fwidth,
height=\fheight,
ybar stacked,bar shift=-20pt,ymin=0,ymax=1.3,
xlabel style={font=\color{white!15!black}},
xlabel={$T_{\rm RT} \text{  [s]}$},
ylabel style={font=\color{white!15!black}},
xlabel style={font=\color{white!15!black}},
ylabel={Rate [Gbps]},
label style={font=\footnotesize},
axis background/.style={fill=white},
title style={font=\bfseries},
xmajorgrids,
ymajorgrids]]

\addplot [ fill=mycolor1,  draw=black,area legend]  
coordinates{(1,1.112) (2,0.6363) (3,0.591) (4,0.554) (5,0.495)};

\addplot [forget plot, fill=mycolor1, draw=black, postaction={pattern=north east lines}, area legend]  
coordinates{(1,0) (2,0) (3,0) (4,0) (5,0)};

\end{axis}

\begin{axis}[
width=\fwidth,
ticks=none,
height=\fheight,
ybar stacked,bar shift=-10pt,ymin=0,ymax=1.3]]  

\addplot+[ fill=mycolor2, draw=black,area legend] 
coordinates{%
(1,0) (2,1.015) (3,0.7738) (4,0.688) (5,0.548) };

\addplot+[ forget plot,fill=white, postaction={pattern=north east lines},area legend] 
coordinates{%
(1,1.161) (2,0) (3,0) (4,0) (5,0) };

\end{axis}

\begin{axis}[
width=\fwidth,
height=\fheight,
ticks=none,
ybar stacked,bar shift=0pt,ymin=0,ymax=1.3]]  

\addplot+[ fill=mycolor3,draw=black, area legend] 
coordinates{%
(1,0) (2,0) (3,1.001) (4,0.726) (5,0.599) };

\addplot+[forget plot, fill=white, postaction={pattern=north east lines},area legend] 
coordinates{%
(1,1.161) (2,1.016) (3,0) (4,0) (5,0) };

\end{axis}

\begin{axis}[
width=\fwidth,
height=\fheight,
ticks=none,
ybar stacked,bar shift=10pt,ymin=0,ymax=1.3]]   

\addplot+[ fill=mycolor4, draw=black,area legend] 
coordinates{%
(1,0) (2,0) (3,0) (4,0.938) (5,0.6963) };

\addplot+[forget plot, fill=white, postaction={pattern=north east lines},area legend] 
coordinates{%
(1,1.161) (2,1.016) (3,1.001) (4,0) (5,0) };

\end{axis}

\begin{axis}[
width=\fwidth,
height=\fheight,
ticks=none,
ybar stacked,bar shift=20pt,ymin=0,ymax=1.3,
legend style={legend cell align=left, align=left, draw=white!15!black, at={(0.5,1.08)},/tikz/every even column/.append style={column sep=0.35cm},
  anchor=north ,legend columns=-1}]]   

\addplot+[forget plot, color=red, fill=white, postaction={pattern=north east lines},area legend] 
coordinates{%
(1,1.161) (2,1.016) (3,1.001) (4,0.938) (5,0) };

\addplot+[ fill=mycolor1, draw=black,area legend] 
coordinates{%
(1,0) (2,0) (3,0) (4,0) (5,0) };

\addplot+[ fill=mycolor2, draw=black,area legend] 
coordinates{%
(1,0) (2,0) (3,0) (4,0) (5,0) };

\addplot+[ fill=mycolor3, draw=black,area legend] 
coordinates{%
(1,0) (2,0) (3,0) (4,0) (5,0)};

\addplot+[ fill=mycolor4, draw=black,area legend] 
coordinates{%
(1,0) (2,0) (3,0) (4,0) (5,0) };

\addplot+[ fill=mycolor5, draw=black,area legend] 
coordinates{%
(1,0) (2,0) (3,0) (4,0) (5,0.8957) };

\legend{$T_{H} = 0.01$ s., $T_{H} = 0.1$ s., $T_{H} = 0.2$ s., $T_{H} = 0.5$ s., $T_{H} = 1$ s.}
\end{axis}

%
%
%
%
%
%
%
%
%
%
%
%
%
%
%
%
%
%

\end{tikzpicture}%
        \caption{\footnotesize Average rate vs. RT periodicity $T_{\rm RT}$, for different values of $T_H$. The mmWave SCell density is kept constant at $M=70$ SCell/km$^2$. White bars are referred to not remarkable cases, since $T_{\rm RT} < T_H$. The user's speed is $v=20$~s. }
  \label{fig:HO1}
   \end{figure}
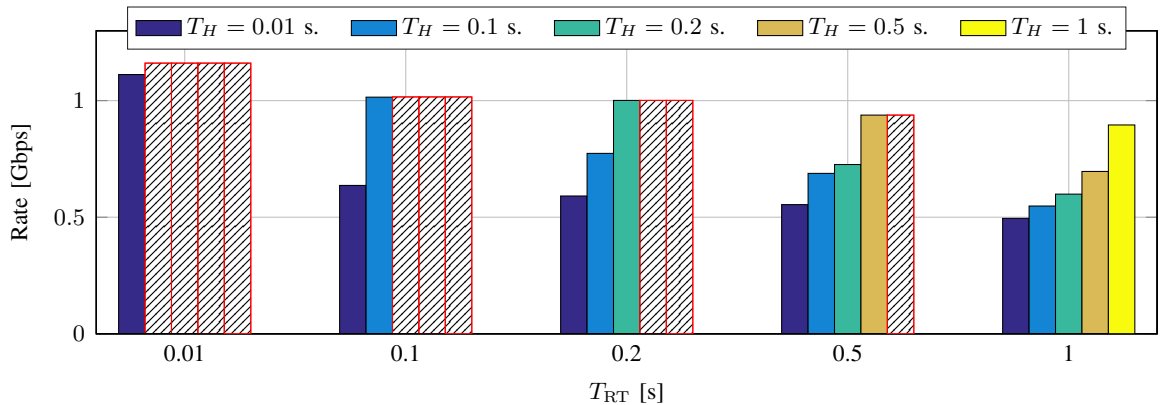

The test user moves at a constant speed $v=20 $ m/s towards a specific direction. Due to its motion and to the variability of the mmWave channel over time, it needs to periodically handover or switch its transmitting beam, to recover a good communication quality.
The large scale fading parameters of the channel are updated every $T_H$ s, while the small scale fading parameters are constantly updated every time slot.
When building a new CRT every $T_{\rm RT}$ s, the MCell can select, by looking at the best saved entry, the new serving SCell for the UE, or just select the new beam pair the transceiver has to set, in order to maximize the communication throughput. 
We just consider the case $T_{\rm RT} \geq T_H$, as otherwise the rate would almost be constant for all values of $T_H$ (since the beam pair would be updated before the channel even changes its large scale fading parameters).

According to Fig. \ref{fig:HO1}, when $T_{\rm RT}$ increases, the average rate decreases, since fewer RTs are exchanged and the beam pair between the user and its serving SCell is monitored less frequently.
This means that, when the channel changes (due to a pathloss condition modification or  to an adaptation of the propagation characteristics) or when the user misaligns with its SCell (due to its motion), the communication quality is not  immediately recovered and the throughput is affected by portions of time where suboptimal  network settings are chosen.
We also observe that, when $T_H$ increases, the average rate also increases since the channel varies less rapidly, so the rate can assume more stable values even if the SCell-UE beam pair is monitored less frequently.
In fact, even if a change in the \textbf{H} matrix's  large scale fading parameters represents the strongest cause for the user's rate slump,  if we consider flat and stable channels, we can accept more rare report tables (and consequently trigger more rare handover and beam switch operations) and still provide  sufficiently good communication quality values.

   \begin{figure}[t!]
     \centering
             \begin{subfigure}[t!]{0.45\textwidth}
             		\setlength{\belowcaptionskip}{0cm}
	\setlength{\belowcaptionskip}{0cm}
	\setlength\fwidth{0.78\columnwidth}
	\setlength\fheight{0.65\columnwidth}
%
%
\definecolor{mycolor1}{rgb}{0.20810,0.16630,0.52920}%
\definecolor{mycolor2}{rgb}{0.21783,0.72504,0.61926}%
\definecolor{mycolor3}{rgb}{0.97630,0.98310,0.05380}%

\pgfplotsset{
tick label style={font=\footnotesize},
label style={font=\footnotesize},
legend  style={font=\footnotesize}
}

\begin{tikzpicture}

\begin{axis}[%
width=\fwidth,
height=\fheight,
at={(0\fwidth,0\fheight)},
scale only axis,
bar shift auto,
xmin=0,
xmax=5,
xtick={1,2,3,4},
xticklabels={{0.01},{0.1},{0.5},{1}},
xlabel style={font=\color{white!15!black}},
xlabel={$\text{T}_{\text{RT}}\text{ [s]}$},
ylabel style={font=\color{white!15!black}},
xlabel style={font=\color{white!15!black}},
ylabel={Rate [Gbps]},
extra y ticks={0.73305},
extra y tick labels={0.75},
extra y tick style={grid style = dashed, color = red},
label style={font=\footnotesize},
ymin=0,
ymax=1.8,
axis background/.style={fill=white},
xmajorgrids,
ymajorgrids,
legend style={legend cell align=left, align=left, draw=white!15!black}
]
\addplot[ybar, bar width=0.18, fill=white!60!orange, draw=black, area legend] table[row sep=crcr] {%
1	0.73305\\
2	0.45647\\
3	0.40964\\
4	0.38564\\
};
\addplot[forget plot, color=white!15!black] table[row sep=crcr] {%
0	0\\
6	0\\
};
\addlegendentry{$M = 30$ SCell/km$^\text{2}$}

\addplot[ybar, bar width=0.18, fill=orange, draw=black, area legend] table[row sep=crcr] {%
1	1.112\\
2	0.63633\\
3	0.554\\
4	0.495\\
};
\addplot[forget plot, color=orange] table[row sep=crcr] {%
0	0\\
6	0\\
};
\addlegendentry{$M = 70$ SCell/km$^\text{2}$}

\addplot[ybar, bar width=0.18, fill=red!60!pink, draw=black, area legend] table[row sep=crcr] {%
1	1.2689\\
2	0.70077\\
3	0.64305\\
4	0.6007\\
};
\addplot[forget plot, color=white!15!black] table[row sep=crcr] {%
0	0\\
6	0\\
};
\addlegendentry{$M = 100$ SCell/km$^\text{2}$}

\end{axis}

\end{tikzpicture}%
    \caption{\footnotesize  Average rate vs. RT periodicity $T_{\rm RT}$. The large scale fading parameters of \textbf{H} are updated every ${T_H = 10}$~ms.}
      \end{subfigure} \qquad \quad
      \begin{subfigure}[t!]{0.45\textwidth}
      		\setlength{\belowcaptionskip}{0cm}
	\setlength{\belowcaptionskip}{0cm}
	\setlength\fwidth{0.78\columnwidth}
	\setlength\fheight{0.65\columnwidth}
%
%
\definecolor{mycolor1}{rgb}{0.20810,0.16630,0.52920}%
\definecolor{mycolor2}{rgb}{0.21783,0.72504,0.61926}%
\definecolor{mycolor3}{rgb}{0.97630,0.98310,0.05380}%

\pgfplotsset{
tick label style={font=\footnotesize},
label style={font=\footnotesize},
legend  style={font=\footnotesize}
}

\begin{tikzpicture}

\begin{axis}[%
width=\fwidth,
height=\fheight,
at={(0\fwidth,0\fheight)},
scale only axis,
bar shift auto,
xmin=0,
xmax=5,
xtick={1,2,3,4},
xticklabels={{0.01},{0.1},{0.5},{1}},
xlabel style={font=\color{white!15!black}},
xlabel={$\text{T}_{\text{RT}}\text{ [s]}$},
ylabel style={font=\color{white!15!black}},
xlabel style={font=\color{white!15!black}},
ylabel={Rate [Gbps]},
label style={font=\footnotesize},
ymin=0,
ymax=1.8,
axis background/.style={fill=white},
xmajorgrids,
ymajorgrids,
legend style={legend cell align=left, align=left, draw=white!15!black}
]
\addplot[ybar, bar width=0.18, fill=white!60!orange, draw=black, area legend] table[row sep=crcr] {%
1	0.7356\\
2	0.63664\\
3	0.52\\
4	0.42162\\
};
\addplot[forget plot, color=white!15!black] table[row sep=crcr] {%
0	0\\
6	0\\
};
\addlegendentry{$M = 30$ SCell/km$^\text{2}$}

\addplot[ybar, bar width=0.18, fill=orange, draw=black, area legend] table[row sep=crcr] {%
1	1.1614\\
2	1.0148\\
3	0.77376\\
4	0.548\\
};
\addplot[forget plot, color=white!15!black] table[row sep=crcr] {%
0	0\\
6	0\\
};
\addlegendentry{$M = 70$ SCell/km$^\text{2}$}

\addplot[ybar, bar width=0.18, fill=red!60!pink, draw=black, area legend] table[row sep=crcr] {%
1	1.2626\\
2	1.1061\\
3	0.87488\\
4	0.6139\\
};
\addplot[forget plot, color=white!15!black] table[row sep=crcr] {%
0	0\\
6	0\\
};
\addlegendentry{$M = 100$ SCell/km$^\text{2}$}

\end{axis}
\end{tikzpicture}%
    \caption{\footnotesize   Average rate vs. RT periodicity $T_{\rm RT}$. The large scale fading parameters of \textbf{H} are updated every ${T_H = 100}$~ms.}
      \end{subfigure}
\caption{Results of the handover and beam tracking simulations, for different SCell densities. The user's speed is $v=20$~/s.}
\label{fig:HO2}
\end{figure}
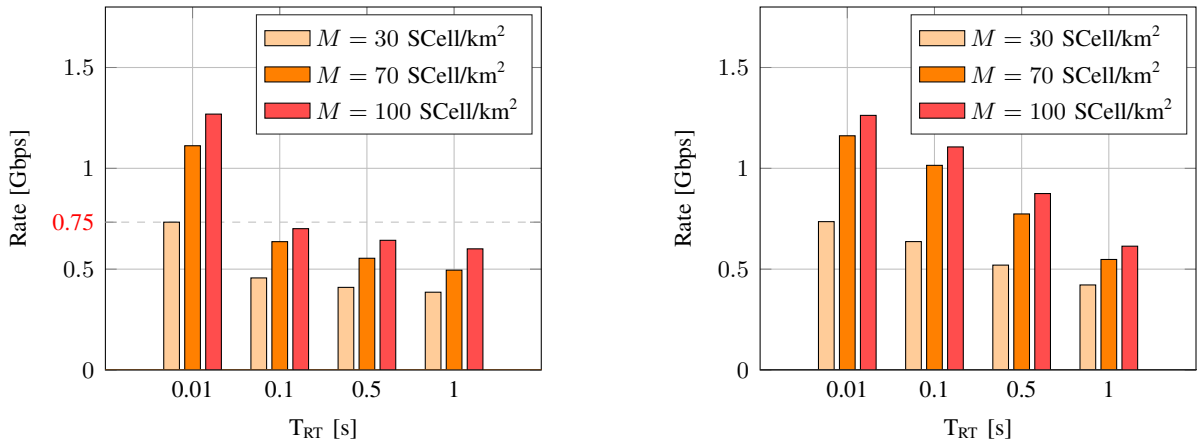

According to Fig. \ref{fig:HO2} and as we pointed our previously,  the  average rate increases for increasing mmWave SCell density values.
Moreover, higher rates are experienced when ${T_H=100}$~ms (Fig. \ref{fig:HO2}(b)), with respect to the $10$ ms case, since the channel changes less rapidly. 
Additionally, if we observe Fig. \ref{fig:HO2}(a), we note how a $0.75$ Gbps rate can be  achieved either with a $30$ SCell/km$^2$ density and $10$ ms $T_{RT}$, or with a $100$ SCell/km$^2$ density and $100$ ms $T_{RT}$: the tradeoff oscillates between infrastructure cost and signaling overhead.

Finally, it is interesting to notice that the main advantage when increasing the cell density is observed from $M=30$~SCell/km$^2$ to $M=70$~SCell/km$^2$. In fact such rate gain reflects the transition from a user outage regime to a LoS/NLoS regime while, as we persistently keep on densifying the network, the deployment of more SCells leads to a considerable increase of the system complexity, while providing a limited increase of the rate.



\subsection{Initial Access Performance}
\label{sec:IA_res}

As assessed in Sec. \ref{sec:comparison_DL}, faster attachment decisions can be made when an uplink multi-connectivity configuration is preferred. 
This result enables a faster initial access scheme too, when digital beamforming is chosen.
We also claim that the use of the supervising 4G-LTE MCell enables a more fair cell selection operation as well. 
Unlike in the traditional approach, a multi-connectivity initial access procedure possibly provides the user two different attachment policies: the UE may thus connect (i) to the SCell guaranteeing the highest SINR (max-SNR rule) or, knowing the current  load of each eNB, (ii) to the  SCell ensuring the highest rate (max-rate~rule).

In order to compare the two presented attachment policies, we use  \emph{Jain's fairness index},  which is used to determine whether users are receiving a fair share of the system resources and are thus experiencing a rate comparable to that of other users in the cellular system. This index is defined as:
\begin{equation}
J = \frac{\left(\sum_{i=1}^N R_i\right)^2}{N \sum_{i=1}^N R_i^2},
\label{Jain}
\end{equation}
 where  $N$ is the number of users in the system and $R_i$ is the rate experienced by the $i$-th user. The result ranges from $1/N$ (worst case) to $1$ (best case), and it is maximum when all users receive the same allocation.

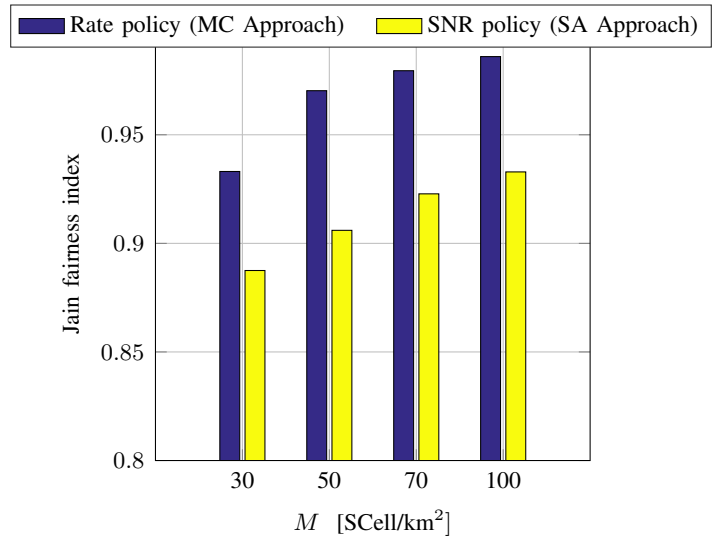
\begin{figure}[t!]
     \centering
     		\setlength{\belowcaptionskip}{0cm}
	\setlength{\belowcaptionskip}{0cm}
	\setlength\fwidth{0.35\textwidth}
	\setlength\fheight{0.35\textwidth}
%
%
\definecolor{mycolor1}{rgb}{0.20810,0.16630,0.52920}%
\definecolor{mycolor2}{rgb}{0.97630,0.98310,0.05380}%
\pgfplotsset{
tick label style={font=\footnotesize},
label style={font=\footnotesize},
legend  style={font=\footnotesize}
}

\begin{tikzpicture}

\begin{axis}[%
width=\fwidth,
height=\fheight,
at={(0\fwidth,0\fheight)},
scale only axis,
bar shift auto,
xmin=0,
xmax=5,
xtick={1,2,3,4},
xticklabels={{30},{50},{70},{100}},
xlabel style={font=\color{white!15!black}},
xlabel={$M$ $ \text{ [SCell/km}^2]$},
ylabel style={font=\color{white!15!black}},
xlabel style={font=\color{white!15!black}},
ymin=0.8,
ymax=1,
ylabel={Jain fairness index},
label style={font=\footnotesize},
axis background/.style={fill=white},
xmajorgrids,
ymajorgrids,
legend style={legend cell align=left, align=left, at={(1.28,1.05)}, draw=white!15!black, /tikz/every even column/.append style={column sep=0.35cm}, legend columns = -1}]

\addplot[ybar, bar width=0.229, fill=mycolor1, draw=black, area legend] table[row sep=crcr] {%
1	0.9331\\
2	0.9703\\
3	0.9795\\
4	0.986\\
};
\addlegendentry{Rate policy  (MC Approach)}

\addplot[ybar, bar width=0.229, fill=mycolor2, draw=black, area legend] table[row sep=crcr] {%
1	0.8875\\
2	0.906\\
3	0.9228\\
4	0.9329\\
};
\addlegendentry{SNR policy (SA Approach)}

\end{axis}
\end{tikzpicture}%
        \caption{\footnotesize Jain's fairness index of the rate vs.  SCell density, for the initial access procedure. Users within an area of radius ${R_C = 70}$~m attach to their best SCell according to a maximum rate or maximum SINR policy. }
        \label{fig:IA_Jain}
 \end{figure}

In Fig. \ref{fig:IA_Jain}, we plot   Jain's fairness index for the rate experienced by users within an area of radius ${R_C = 70}$ meters, when attaching  either according to the max-SINR rule (as in traditional schemes) or the max-rate rule (by exploiting the MC procedure).
As expected, this last attachment policy provides  higher fairness to the network: asymptotically,  a UE accessing the network at time $t$ will likely find all the SCells in the same load conditions, guaranteeing comparable rates.
On the other hand, by following a max-SINR attachment policy, users will tend to connect to the same SCells showing the instantaneous highest signal strengths (and thus overloading them), and avoiding instead nodes that provide lower SINR values (but possibly higher rates, due to their low traffic loads).

We finally notice that  Jain's fairness index in Fig. \ref{fig:IA_Jain} increases with $M$ for both schemes. In fact, when densifying the network, the SCells  ensure more similar propagation conditions to the users, which in turn experience more balanced SINR (and rate) values.

\subsection{RLF Recovery Performance}
\label{sec:RLF_res}


According to the scenario described in Sec. \ref{sec:RLF}, we define  $R$ as the optimal rate experienced when no obstacles affect the signal propagation, and $r$ as the suboptimal rate experienced when a suboptimal backup beam pair is selected, after the primary path is~obstructed.

Assume that a blockage event is detected at time ${T_{\rm arr}\sim \mathcal{U}(0,T_{\rm RT}) = T_{\rm RT} p}$, with ${p\in(0,1)}$, and lasts for $T_B$ s.
We aim at finding the \emph{rate gain} ($R_G$), namely the ratio between the rate experienced when the MC procedure is employed to establish a backup beam pair between the user and its serving SCell after a blockage is detected ($R_{\rm WB}$), and the rate perceived when no actions are taken ($R_{\rm OB}$).

We focus on the situation in which the obstacle is no longer present when the new CRT is generated (${T_{\rm RT} \geq 2T_B}$), otherwise the beam pair would be updated when the obstacle is still obstructing the best path, thus still reducing the average rate.
Then, the rate $R_{\rm WB}$ experienced when reacting after the blockage is detected by selecting a suboptimal solution in the RT instances can be computed (for a fixed time window $T_{\rm RT}$), as:
\begin{equation}
\begin{split}
R_{\rm WB} &= \frac{R T_{\rm arr} + r T_B + R(T_{\rm RT}-T_{\rm arr}-T_B)}{T_{\rm RT}} = \dots = \frac{R (T_{\rm RT} - T_B) + rT_B}{T_{\rm RT}} \\
\end{split}
\end{equation}

If no actions are taken, after the obstacle has been detected, the rate $R_{\rm OB}$ is:
\begin{equation}
\begin{split}
R_{\rm OB} &= \frac{R T_{\rm arr} + 0 T_B + R(T_{\rm RT}-T_{\rm arr}-T_B)}{T_{\rm RT}} = \dots = \frac{R (T_{\rm RT} - T_B) }{T_{\rm RT}}
\end{split}
\end{equation}

The average rate gain ($R_G$) between the two options is:

\begin{equation}
R_G = \frac{R_{\rm WB}}{R_{\rm OB}} = 1 + \frac{r}{R} \cdot \frac{T_B}{T_{\rm RT}-T_B}
\label{eq:R_G}
\end{equation}


     \begin{figure}
     \centering
     		\setlength{\belowcaptionskip}{0cm}
	\setlength{\belowcaptionskip}{0cm}
	\setlength\fwidth{0.4\textwidth}
	\setlength\fheight{0.35\textwidth}
	\input{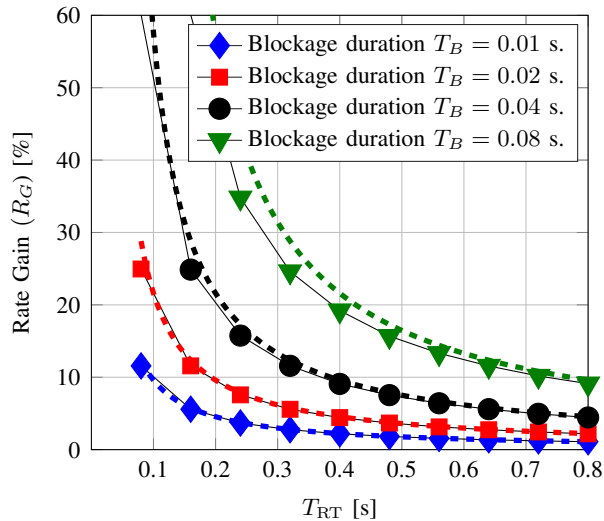}
        \caption{\footnotesize Rate gain experienced when applying a backup procedure for the RLF recovery vs. RT periodicity $T_{\rm RT}$, for different blockage scenarios. The obstacle duration is $T_B$ s and is detected after $T_{\rm arr}$ s.}
        \label{fig:RLF_res}
   \end{figure}

In Fig. \ref{fig:RLF_res}, we first notice that $R_G > 0$ for all values of  $T_{\rm RT}$ and $T_B$, making it clear that having a second available link (in case the primary one is blocked) guarantees improved communication throughput performance with respect to a traditional scheme in which a backup configuration is not available.
Furthermore, when $T_{\rm RT}$ is sufficiently large, so when $T_{RT} \gg 2T_B$, the simulation curves asymptotically  overlap with the dashed lines  plotting   Eq. \eqref{eq:R_G}.
Fig. \ref{fig:RLF_res} shows also that, for a fixed blockage duration $T_B$, as $T_{\rm RT}$ increases, the rate gain $R_G$ decreases. In fact the portion of time in which the user would experience zero gain (if no actions are taken when the primary path is obstructed) proportionally decreases within the time window of length $T_{\rm RT}$, making it less convenient to select a backup beam pair to overcome the blockage issue. 

Finally, we see that, when $T_B$ increases, the rate gain $R_G$ increases as well, due to the increased enhancement provided by the use of a suboptimal beam pair after a blockage event occurs, with respect to the baseline algorithm in which no actions are taken till the reception of a new CRT.

\subsection{Final Comments}
To sum up, a comparison between the uplink multi-connectivity framework presented in this work and a traditional downlink standalone approach has been made. 
Specifically, we concluded that a MC scheme:
\begin{itemize}
\item[(i)] offers significantly reduced access delays when a digital beamforming  architecture is~chosen;
\item[(ii)] leveraging on the LTE eNB to deal with outage events, guarantees higher average data rates to the system, especially when considering sparse environments;
\item[(iii)] enables an energy-efficient mobility management scheme for the mobile terminal, the most energy-constrained entity in the cellular network;
\item[(iv)] stabilizes the rate and flattens most of the variations and fluctuations a mmWave channel is usually affected by, improving the performance of real-time applications requiring a long-term stable throughput.
\end{itemize}

Furthermore, we proved that the proposed framework enables the design of efficient 5G control plane applications that specify how a user should attach to the network and maintain its connectivity. In particular, we showed that a multi-connectivity approach:
\begin{itemize}
\item[(i)] enables performing mobility management operations even when considering highly dynamic environments;
\item[(ii)] enables fast and  fair initial access operations,~if~a~max-rate~attachment~policy~is~chosen;
\item[(iii)] guarantees an efficient radio-link failure recovery when a backup steering direction is set, in case the primary path is obstructed.
\end{itemize}

\section{Conclusions and Future Work}
\label{concl}

A  challenge for the feasibility of a 5G mmWave system is the high susceptibility to the rapid channel dynamics that affect a mmWave environment.
In order to deal with these channel variations, a periodic directional sweep should be performed, to constantly monitor the directions of transmission of each potential link and to adapt the beam steering when a power signal drop is detected.
In this work, we have presented a  measurement reporting system that allows a supervising centralized entity, such as  a base station operating in the legacy band, to periodically collect multiple reports on the overall channel propagation conditions, to enable efficient scheduling  and mobility management decisions.
We argue that the proposed uplink multi-connectivity approach enables  more rapid,  robust, performing and energy efficient network operations at the mobile terminal, in particular when considering very unstable channels and highly populated systems.
Moreover, we proved that fast and fair initial user association, enhanced handover management and reactive radio-link failure recovery can be guaranteed when a multi-connectivity configuration is preferred.

As part of our future work, we will design control applications that monitor and keep memory  of the received signal strength variance, to better capture the dynamics of the channel and bias the cell selection strategy of delay-sensitive applications towards more robust cells.

\bibliographystyle{IEEEtran}
\bibliography{bibliography}

\end{document}